\documentclass[aps,prxquantum,twocolumn,notitlepage]{revtex4-1}
\usepackage{amsmath,amssymb}	
\usepackage{natbib}
\usepackage{url}
\usepackage{bm} 
\usepackage{mathtools}
\usepackage{braket}
\usepackage{hyperref}
\usepackage{graphicx}
\usepackage{color}
\usepackage{xcolor}
\usepackage[caption=false]{subfig}
\usepackage{epstopdf}
\usepackage{array}
\usepackage{empheq}
\usepackage{makecell}
\usepackage{booktabs}

\DeclareMathOperator{\Tr}{Tr}

\newcommand{\HO}{\hat{\mathcal{H}}}

\begin{document}
\title{Mitigating off-resonant error in the cross-resonance gate}
\author{Moein Malekakhlagh} \email{Electronic address: moein.malekakhlagh@ibm.com}
\author{Easwar Magesan} \email{Electronic address: emagesa@us.ibm.com}
\affiliation{IBM Quantum, IBM Thomas J. Watson Research Center, 1101 Kitchawan Rd, Yorktown Heights, NY 10598}
\date{\today}
\begin{abstract}
Off-resonant error for a driven quantum system refers to interactions due to the input drives having non-zero spectral overlap with unwanted system transitions. For the cross-resonance gate, this includes leakage as well as off-diagonal computational interactions that lead to bit-flip error on the control qubit. In this work, we quantify off-resonant error, with more focus on the less studied off-diagonal control interactions, for a \textit{direct} CNOT gate implementation. Our results are based on numerical simulation of the dynamics, while we demonstrate the connection to \textit{time-dependent} Schrieffer-Wolff and Magnus perturbation theories. We present two methods for suppressing such error terms. First, pulse parameters need to be optimized so that off-resonant transition frequencies coincide with the local minima due to the pulse spectrum sidebands. Second, we show the advantage of a $Y$-DRAG pulse on the control qubit in mitigating off-resonant error. Depending on qubit-qubit detuning, the proposed methods can improve the average off-resonant error from \textit{approximately} $10^{-3}$ closer to the $10^{-4}$ level for a direct CNOT calibration.
\end{abstract}
\maketitle	

\section{Introduction}
\label{Sec:Intro}

Cross-resonance (CR) is a microwave-activated two-qubit gate performed by driving one of the qubits (control) at the frequency of the other (target) \cite{Paraoanu_Microwave_2006, Rigetti_Fully_2010}. In this architecture, superconducting qubits \cite{Nakamura_Coherent_1999, Wallraff_Strong_2004, Koch_Charge_2007, Clarke_Superconducting_2008}, typically fixed-frequency transmons \cite{Koch_Charge_2007}, connect via a mediating bus resonator, resulting in a static qubit-qubit exchange interaction. The CR protocol induces various two-qubit interactions \cite{Sheldon_Procedure_2016, Magesan_Effective_2020, Kirchhoff_Optimized_2018, Tripathi_Operation_2019, Malekakhlagh_First-Principles_2020, Sundaresan_Reducing_2020, Heya_Cross_2021}, with $ZX$ as the dominant rate, through which a CNOT gate can be calibrated. Simplicity in implementation, resilience to charge and flux noise, and scalability has made CR architecture the leading workhorse for current IBM quantum processors \cite{Cross_Validating_2019, Sundaresan_Reducing_2020, Jurcevic_Demonstration_2021}. 	

Improving CR gate fidelity necessitates both an accurate understanding of the effective interactions as well as precise microwave control. In particular, a multi-level analysis of the dynamics \cite{Sheldon_Procedure_2016, Magesan_Effective_2020, Kirchhoff_Optimized_2018, Tripathi_Operation_2019, Malekakhlagh_First-Principles_2020, Sundaresan_Reducing_2020} is required for CR gate implementation with weakly anharmonic transmon qubits \cite{Koch_Charge_2007}. Higher qubit states can lead to both renormalization of interactions in the computational subspace \cite{Magesan_Effective_2020, Malekakhlagh_First-Principles_2020, Sundaresan_Reducing_2020} as well as out-of-computational leakage \cite{Wood_Quantification_2018, Tripathi_Operation_2019}. Generally, to optimize the \textit{coherent} fidelity, we need to maximize the desired $ZX$ rate and minimize unwanted computational and leakage interactions.

There are two \textit{main} CNOT calibration schemes based on CR architecture. First, an \textit{echo} sequence consisting of \textit{two} CR pulses with flipped amplitude accompanied with single-qubit rotations \cite{Sheldon_Procedure_2016, Malekakhlagh_First-Principles_2020, Sundaresan_Reducing_2020, Jurcevic_Demonstration_2021}. The echo removes certain unwanted rates such as the $ZI$, $ZZ$ and $IX$, while induces higher order $IY$ and $IZ$ error terms \cite{Malekakhlagh_First-Principles_2020, Sundaresan_Reducing_2020}. Reference~\cite{Jurcevic_Demonstration_2021} demonstrated a $280$ ns echoed CR gate with an average fidelity of $99.40\%$. Second, a \textit{direct} CNOT calibration with a single CR pulse \cite{Tripathi_Operation_2019, Kandala_Demonstration_2020, Jurcevic_Demonstration_2021}. To this aim, we want no operation on the target qubit when the control is in state $\ket{0_c}$, hence canceling $IX+ZX$ via a separate drive on the target, and a $\pi$ rotation on the target when the control is in state $\ket{1_c}$. Therefore, the $IX$ rate is \textit{not} an error term anymore, and CR gate speed is determined by \textit{twice} the $ZX$ ($IX$) rate, resulting in a faster gate. Equipped with multiple-path interference couplers \cite{Mundada_Suppression_2019} to suppress the static $ZZ$ rate, and using virtual frame change \cite{McKay_Efficient_2017} to cancel out Stark shifts in software, Ref.~\cite{Kandala_Demonstration_2020} demonstrated a $180$ ns gate with $99.77\%$ average gate fidelity.

Schrieffer-Wolff Perturbation Theory (SWPT) \cite{Schrieffer_Relation_1966, Boissonneault_Dispersive_2009, Bravyi_Schrieffer_2011, Gambetta_Analytic_2011, Malekakhlagh_Lifetime_2020, Petrescu_Lifetime_2020, Magesan_Effective_2020, Malekakhlagh_First-Principles_2020, Petrescu_Accurate_2021} is a central method in our analytical understanding of \textit{effective} interactions. SWPT provides effective models by averaging high-frequency off-resonant processes systematically. Through a series of perturbative frame transformations, the interactions are partitioned into resonant (effective) and off-resonant categories. The effective interactions come from processes that connect states with equal frequency in the rotating-frame of the drive, while off-resonant interactions have a net non-zero transition frequency. In contrast to rotating-wave approximation (RWA), which simply discards off-resonant terms, SWPT takes them into account by solving for and storing the relevant frame transformations. Hence, contributions that are not resonant at a specific order may lead to resonant interactions via non-trivial higher order mixings. The drive scheme, and the corresponding energy diagram, determines the choice of the effective frame. For CR, where the drive is resonant with the target qubit, the effective SWPT frame is block-diagonal (BD) with respect to the control \cite{Magesan_Effective_2020}. Under the BD approximation, analytical estimates for CR gate parameters have been derived in Refs.~\cite{Magesan_Effective_2020, Malekakhlagh_First-Principles_2020} for \textit{constant-amplitude} continuous wave (CW) drive.  

\begin{figure}[t!]
\includegraphics[scale=0.51]{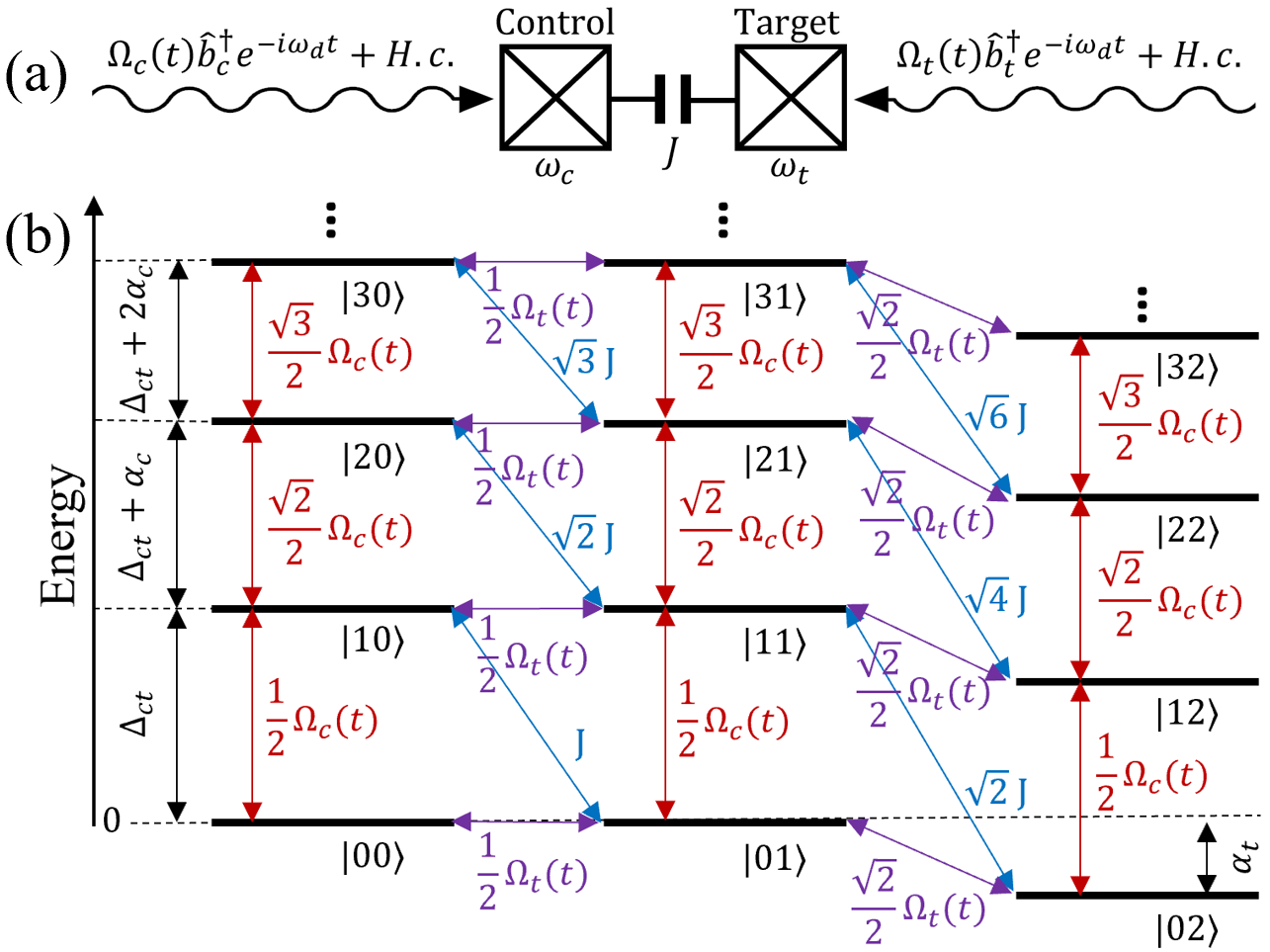}
\centering
\caption{(a) CR gate schematic with separate drives on each qubit. The drive amplitudes $\Omega_{j}(t)\equiv \Omega_{jx}(t)+i\Omega_{jy}(t)$, $j=c,t$ are complex-valued to allow for more involved control schemes like DRAG \cite{Motzoi_Simple_2009, Chow_Optimized_2010, Gambetta_Analytic_2011, Schutjens_Single-Qubit_2013}. The drive frequency is set to the target qubit frequency. (b) Energy diagram in the rotating frame of the drive under RWA (see Sec.~\ref{Sec:DirCNOT}). The control-target detuning is denoted by $\Delta_{ct}\equiv\omega_c-\omega_t$, and control and target anharmonicities by $\alpha_{c}$ and $\alpha_{t}$, respectively.}
\label{fig:CRSchematicPlusEnergyDiagram}
\end{figure}

In this paper, we study off-resonant error due to interactions that originate from the CR drive frequency being detuned from states of the control qubit. The corresponding dominant unwanted transitions are $\ket{0_c}\leftrightarrow\ket{1_c}$, $\ket{0_c}\leftrightarrow\ket{2_c}$ and $\ket{1_c}\leftrightarrow\ket{2_c}$ (see Fig.~\ref{fig:CRSchematicPlusEnergyDiagram}). Although off-resonant error is present for a CW drive, there is an intricate interplay with the pulse shape and in particular the pulse ramps. To model this, we employ numerical simulations based on Magnus expansion \cite{Magnus_Exponential_1954, Blanes_Magnus_2009, Blanes_Pedagogical_2010, Hairer_Geometric_2006}. Furthermore, we extend our SWPT formalism for CR in Refs.~\cite{Magesan_Effective_2020, Malekakhlagh_First-Principles_2020} to the time-dependent case and make a connection to the Magnus method. In particular, the main role of \textit{time-independent} and \textit{dependent} perturbations are to account for \textit{how strong} the drive amplitude and \textit{how fast} (non-adiabatic) the pulse ramps are compared to the system transition frequencies, respectively. These two effects are independent in general, however, in the context of gate calibration, they become related based on a fixed rotation angle imposed by the intended gate. 

Generally, to mitigate off-resonant error, CR drive should have minimal spectral content at the unwanted transition frequencies. Qubit-qubit detuning and anharmonicity determine the relative configuration of off-resonant transitions in the rotating-frame of the drive \cite{Tripathi_Operation_2019, Malekakhlagh_First-Principles_2020}, where the error due to one or sometimes multiple transitions can be noticeable. We take the standard square Gaussian pulse, i.e. flat top with Gaussian ramps, and demonstrate further improvement. The most immediate refinement comes from optimization of the pulse rise time known as Gaussian shaping \cite{Chow_Optimized_2010, Gambetta_Analytic_2011}. It should be set so that the transition for the most dominant error type overlaps with one of the local minima of the pulse sidebands. Moreover, we show additional improvement by a $Y$-DRAG \cite{Motzoi_Simple_2009, Chow_Optimized_2010, Gambetta_Analytic_2011, Schutjens_Single-Qubit_2013} pulse on the control qubit. We argue that, in essence, DRAG acts as an effective filter that can be tuned to notch the error due to specific off-resonant transitions. 

The remainder of this paper is organized as follows. In Sec.~\ref{Sec:DirCNOT}, we introduce our model for a direct CNOT gate implementation \cite{Tripathi_Operation_2019, Kandala_Demonstration_2020}. In Sec.~\ref{Sec:OffRes}, we discuss the theory behind off-resonant error, by relating time-dependent SWPT to the Magnus method, which extends our earlier results in Refs.~\cite{Magesan_Effective_2020, Malekakhlagh_First-Principles_2020}. Furthermore, using numerical simulation based on Magnus, we quantify the dependence of off-resonant error on pulse parameters. In Sec.~\ref{Sec:DRAG}, we demonstrate the advantage of $Y$-DRAG pulse on the control qubit in suppressing off-resonant error.

There are five appendices. In Appendix~\ref{App:SWPT}, we discuss the derivation of a generalized time-dependent SWPT to account for the underlying pulse shapes. Using the SWPT formalism, Appendix~\ref{App:EffCRHam} derives effective \textit{time-dependent} Hamiltonian rates generalizing Refs.~\cite{Magesan_Effective_2020, Tripathi_Operation_2019, Malekakhlagh_First-Principles_2020}. Appendices~\ref{App:EffU} and~\ref{App:NonBDU} provide the effective time evolution operator and the corresponding leading order non-BD contributions, respectively. In Appendix~\ref{App:OffResQu}, we derive leading order estimates for dominant off-resonant error types, and corresponding DRAG conditions for their suppression.

\section{Direct CNOT}
\label{Sec:DirCNOT}

We consider two coupled transmon qubits \cite{Koch_Charge_2007} with individual drive on each qubit. The transmon Hamiltonian can be approximated in terms of multi-level Kerr oscillators as 
\begin{align}
\HO_q \equiv \sum\limits_{j=c,t} \left[\omega_j \hat{b}_j^{\dag}\hat{b}_j+\frac{\alpha_j}{2}\hat{b}_j^{\dag}\hat{b}_j^{\dag}\hat{b}_j\hat{b}_j \right] \;,
\label{eqn:DirCNOT-Def of H0}
\end{align}
with $\omega_{c,t}$ and $\alpha_{c,t}$ denoting the corresponding frequency and anharmonicity for the control and the target qubits, respectively. The transmon-transmon exchange interaction takes the approximate form 
\begin{align}
\HO_J \equiv J\left( \hat{b}_c^{\dag} \hat{b}_t + \hat{b}_c\hat{b}_t^{\dag} \right) \;,
\label{eqn:DirCNOT-Def of HJ}
\end{align}
where $J$ is the effective exchange rate as a result of either direct capacitive coupling or mediated coupling through a common bus resonator \cite{Magesan_Effective_2020, Malekakhlagh_First-Principles_2020}. Furthermore, we consider separate drives on each qubit as 
\begin{align}
\begin{split}
\HO_{d} (t)  &\equiv \frac{1}{2}\left[\Omega_c^*(t)\hat{b}_c e^{i \omega_d t}+\Omega_c(t) \hat{b}_c^{\dag} e^{-i\omega_d t} \right]\\
&+\frac{1}{2} \left[ \Omega_t^*(t)\hat{b}_t e^{i \omega_d t}+ \Omega_t(t)\hat{b}_t^{\dag} e^{-i \omega_d t}\right],
\label{eqn:DirCNOT-Def of Hd(t)}
\end{split}
\end{align}
with frequency $\omega_d$, same as target qubit frequency, and time-dependent complex-valued pulse amplitudes $\Omega_c(t)\equiv \Omega_{cx}(t)+i\Omega_{cy}(t)$ and $\Omega_t(t)\equiv \Omega_{tx}(t)+i\Omega_{ty}(t)$ (see Fig.~\ref{fig:CRSchematicPlusEnergyDiagram}). The \textit{phase} of the microwave drive determines the axis, in the $X$--$Y$ plane of the target qubit, to which the drive couples. We set the main CR drive to couple to the $X$ quadrature of the target, while the $Y$ axis can be used for DRAG following the same conventions as single-qubit gates \cite{Motzoi_Simple_2009, Gambetta_Analytic_2011, Schutjens_Single-Qubit_2013}. In our numerical and perturbative analysis, the system Hamiltonian is the static part $\HO_s \equiv \HO_q+\HO_J$, accounting for the dressing due to J, and the interaction-frame Hamiltonian is defined as $\HO_I(t)\equiv \exp(i\HO_s t) \HO_d(t) \exp(-i\HO_s t)$.

In Eqs.~(\ref{eqn:DirCNOT-Def of H0})--(\ref{eqn:DirCNOT-Def of Hd(t)}), we have adopted a Kerr model for the qubits, and applied RWA in both drive and exchange interactions. This RWA neglects terms that oscillate at approximately twice the target qubit frequency, of the order of 10 GHz, and does \textit{not} change the physics of the CR gate qualitatively. Our goal here is to work with the simplest model to focus mainly on more dominant error that comes from off-resonant transitions of the control qubit, with transition frequencies of the order of 100 MHz, and study the dependence on pulse parameters. A more precise model for the CR gate was discussed in our earlier work~\cite{Malekakhlagh_First-Principles_2020}, accounting also for eigenstate renormalization due to counter-rotating terms in the Josephson nonlinearity. 

Tuning a direct CNOT gate \cite{Kandala_Demonstration_2020} requires
identity operation on the target qubit when the control is in state $\ket{0_c}$ so that 
\begin{align}
\omega_{ix}(t)+\omega_{zx}(t)=0 \;,
\label{eqn:DirCNOT-Cond IX+ZX=0}
\end{align}
and a $\pi$ rotation around the $X$ axis when the control is in state $\ket{1_c}$ as 
\begin{align}
&\int_{0}^{\tau_p} dt \, [\omega_{ix}(t)-\omega_{zx}(t)]=\pi  \;,
\label{eqn:DirCNOT-Cond IX-ZX=Pi}
\end{align}
where $\omega_{\sigma_j\sigma_k}(t) \equiv (1/2) \text{Tr}\{\HO_{\text{CR,eff}}(t)\left(\hat{\sigma}_j\otimes\hat{\sigma}_k\right)\}$, $j,k\in \{i,x,y,z\}$, are effective Hamiltonian rates defined over the dressed two-qubit pauli operators. For notation simplicity, we use $\hat{I}\equiv\hat{\sigma}_i$, $\hat{X}\equiv\hat{\sigma}_x$, $\hat{Y}\equiv \hat{\sigma}_{y}$, $\hat{Z}\equiv \hat{\sigma}_{z}$ and drop explicit tensor product for two-qubit Pauli operators, e.g. $\hat{Z}\hat{X}\equiv \hat{\sigma}_{z}\otimes \hat{\sigma}_x$. Time-dependent SWPT provides reasonable ballparks for the effective gate parameters (see Appendices~\ref{App:SWPT} and~\ref{App:EffCRHam}). In particular, based on Eq.~(\ref{eqn:DirCNOT-Cond IX+ZX=0}), the main cancellation tone on the target is found as 
\begin{align}
\Omega_{tx} (t)=\frac{J}{\Delta_{ct}} \Omega_{cx}(t)-\frac{\alpha_c J}{2\Delta_{ct}^3 (2\Delta_{ct}+\alpha_c)}\Omega_{cx}^3(t)+O\left(\Omega_{cx}^5\right),
\label{eqn:DirCNOT-PertCond IX+ZX=0}
\end{align}
while for CNOT calibration [Eq.~(\ref{eqn:DirCNOT-Cond IX-ZX=Pi})] the main pulse should satisfy approximately
\begin{align}
\begin{split}
&\frac{2\alpha_c J}{\Delta_{ct}(\Delta_{ct}+\alpha_c)}\Omega_{cx}[\tau_p-s_1(\tau_r)]\\
&-\frac{(9\Delta_{ct}^3+15\Delta_{ct}^2\alpha_c+11\Delta_{ct}\alpha_c^2+3\alpha_c^3)\alpha_c^2 J}{\Delta_{ct}^3(\Delta_{ct}+\alpha_c)^3(2\Delta_{ct}+\alpha_c)(2\Delta_{ct}+3\alpha_c)}\\
&\times \Omega_{cx}^3[\tau_p-s_3(\tau_r)]+O\left(\Omega_{cx}^5\right)=\pi \;.
\end{split}
\label{eqn:DirCNOT-PertCond IX-ZX=Pi}
\end{align}  
Equations~(\ref{eqn:DirCNOT-PertCond IX+ZX=0})--(\ref{eqn:DirCNOT-PertCond IX-ZX=Pi}) provide a useful initial guess for drive parameters and facilitate more involved numerical optimization. We also find non-adiabatic corrections to the effective gate parameters in terms of the pulse derivatives, e.g. terms proportional to $J\dot{\Omega}_{cx}^2(t)\Omega_{cx}(t)$ in $\omega_{ix}(t)$ and $\omega_{zx}(t)$ (see Appendix~\ref{App:EffCRHam}). Here, $\tau_{p}$ and $\tau_{r}$ denote the gate and the rise times, respectively. Moreover, $s_n(\tau_r)$ characterizes the \textit{reduced} area under the curve during the ramps, compared to a square pulse, for the $n$th power of the pulse shape. In our simulations, we use the common square Gaussian pulse \cite{Jurcevic_Demonstration_2021, Kandala_Demonstration_2020}. 

Coherent error, compared to an ideal CNOT, can be traced back to the two aforementioned categories of interactions. Resonant (effective) error arises from faulty resonant rotations, terms like $IX \ (IY)$ and $ZX \ (ZY)$, or static and dynamic frequency shifts of the qubits, like $IZ$, $ZI$ and $ZZ$ rates. Static $ZZ$ can in principle be substantially suppressed via tunable \cite{VanDerPloeg_Controllable_2007, Chen_Qubit_2014, McKay_High-Contrast_2015, Stehlik_Tunable_2021}, opposite-anharmonicity \cite{Ku_Suppression_2020, Zhao_High_2020} and multiple-path \cite{Yan_Tunable_2018, Mundada_Suppression_2019, Sung_Realization_2020, Collodo_Implementation_2020, Xu_High-Fidelity_2020, Kandala_Demonstration_2020} couplers, and also via auxiliary AC Stark tones (siZZle) on the qubits \cite{Wei_Quantum_2021, Mitchell_Hardware_2021}. Furthermore, Stark shifts can be removed effectively through virtual frame change in software \cite{McKay_Efficient_2017, Kandala_Demonstration_2020}. In the following, we explore the physics of off-resonant error. 

\begin{figure}
\includegraphics[scale=0.48]{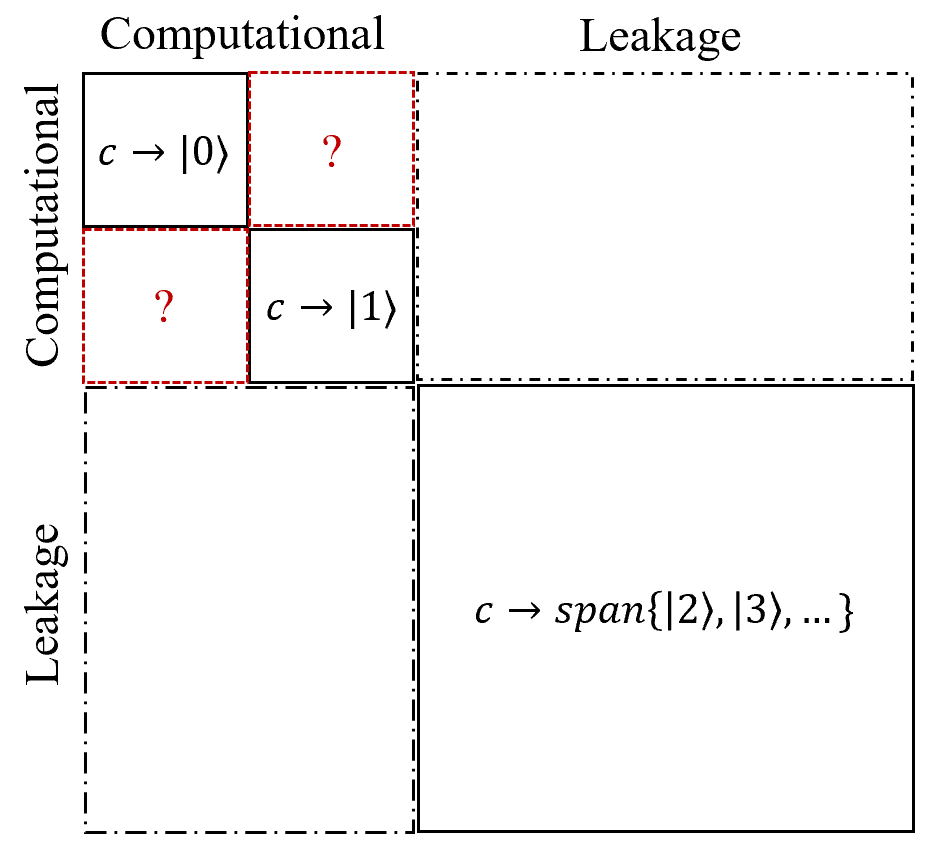}
\centering
\caption{Schematic of the BD subspace shown in solid black, subspace corresponding to coupling between computational and leakage in dashed-dotted black, and non-BD computational subspace in red dotted lines. We are particularly interested in characterizing the error due to non-BD interactions of the form $\hat{\sigma}_x\otimes \hat{\sigma}_k$ or $\hat{\sigma}_y\otimes \hat{\sigma}_k$ for $k=i,x,y,z$.}
\label{fig:OffRes-BDFrame}
\end{figure}
\begin{table*}
\begin{tabular}{|c|c|c|c|}
\hline
Probability & Type & Time-domain & Frequency-domain\\
\hline\hline
$\ket{0_c}\rightarrow \ket{1_c}$ & (1) non-BD & $\approx\frac{1}{4}\Big|\int_{0}^{\tau_p}dt' \Omega_c(t')e^{i\Delta_{ct}t'}\Big|^2$ & $\approx\frac{1}{4}\Big|\int\limits_{-\infty}^{+\infty} \frac{d\omega'}{2\pi} \frac{\tilde{\Omega}_c(\omega')}{\omega'+\Delta_{ct}}\left[e^{i(\omega'+\Delta_{ct})\tau_p}-1\right]\Big|^2$ \\
\hline
$\ket{0_c}\rightarrow \ket{2_c}$ & (2) Leakage & \makecell{$\approx\frac{1}{32}\Big|\int_{0}^{\tau_p}dt'\int_{0}^{t'}dt'' \Omega_c(t')\Omega_c(t'')$ \\ $
\left[e^{i(\Delta_{ct}+\alpha_c)t'}e^{i\Delta_{ct}t''}-e^{i\Delta_{ct}t'}e^{i(\Delta_{ct}+\alpha_c)t''}\right]\Big|^2$} & \makecell{$\approx\frac{1}{32}\Big|\int\limits_{-\infty}^{+\infty}\int\limits_{-\infty}^{+\infty} \frac{d\omega'}{2\pi} \frac{d\omega''}{2\pi}\frac{\alpha_c \tilde{\Omega}_c(\omega') \tilde{\Omega}_c(\omega'')}{(\omega''+\Delta_{ct}+\alpha_c)(\omega''+\Delta_{ct})}$\\
$\frac{\Big[e^{i(\omega'+\omega''+2\Delta_{ct}+\alpha_c)\tau_p}-1\Big]}{\omega'+\omega''+2\Delta_{ct}+\alpha_c}\Big|^2$}\\
\hline
$\ket{1_c}\rightarrow \ket{2_c}$ & (3) Leakage & $\approx\frac{1}{2}\Big|\int_{0}^{\tau_p} dt' \Omega_c(t')e^{i(\Delta_{ct}+\alpha_c)t'}\Big|^2$ & $\approx\frac{1}{2}\Big|\int\limits_{-\infty}^{+\infty} \frac{d\omega'}{2\pi} \frac{\tilde{\Omega}_c(\omega')}{\omega'+\Delta_{ct}+\alpha_c}\Big[e^{i(\omega'+\Delta_{ct}+\alpha_c]\tau_p}-1\Big]\Big|^2$ \\
\hline
\end{tabular}
\caption{Summary of dominant off-resonant error types and leading order overlap integrals in time and frequency domains. The estimates are based on a reduced model of an off-resonantly driven transmon qubit equivalent to vertical ladders in Fig.~\ref{fig:CRSchematicPlusEnergyDiagram} (see also Appendix~\ref{App:OffResQu}). In particular, single-photon transitions are enhanced when there is large overlap with the sideband photons of the pulse, i.e. at $\omega'=-\Delta_{ct}$ and $\omega'=-(\Delta_{ct}+\alpha_c)$ for type 1 and type 3, respectively. Two-photon transition is enhanced due to simultaneous overlap with two sideband photons at $\omega'+\omega''=-(2\Delta_{ct}+\alpha_c)$. Proximity to frequency collisions at $\Delta_{ct}=0,\ -\alpha_c/2,\ -\alpha_c$ enhances the corresponding overlap as the transitions are excited by closer-to-center and hence stronger spectrum sidebands. For strong drive amplitude, comparable to the transition frequencies, more precise estimates can be obtained by \textit{exponentiation} of the Magnus generator, compared to the perturbative expansion used in this table.}
\label{tab:OffResErrSummary}
\end{table*}
           
\section{Off-resonant error}
\label{Sec:OffRes}

Generally, a non-zero overlap of the drive spectrum with an unwanted off-resonant system transition can cause error. For a direct CNOT calibration, the dominant off-resonant error types are bit-flip and leakage on the control qubit \cite{Tripathi_Operation_2019} (see Fig.~\ref{fig:OffRes-BDFrame}).  Unlike resonant error, which can be approximately modeled via time-independent methods accounting only for constant amplitude drive \cite{Magesan_Effective_2020, Tripathi_Operation_2019, Malekakhlagh_First-Principles_2020}, off-resonant error exhibits subtle interplay with the pulse shapes and requires more involved time-dependent methods, such as generalized time-dependent SWPT \cite{Gambetta_Analytic_2011, Malekakhlagh_First-Principles_2020} and Magnus \cite{Magnus_Exponential_1954, Blanes_Magnus_2009, Blanes_Pedagogical_2010}, discussed in the following. 

\subsection{Theory}
\label{SubSec:OffResTheory}

The time evolution operator for the CR gate is found formally as
\begin{align}
\hat{U}_{I} (\tau_p,0) \equiv \mathbb{T} \exp \left[-i \int_{0}^{\tau_p} dt' \HO_I(t') \right] \;,
\label{eqn:OffRes-Def of Ui(t,t0)}
\end{align}
where $\HO_I(t)$ is the drive Hamiltonian~(\ref{eqn:DirCNOT-Def of Hd(t)}), expressed in the interaction frame with respect to $\HO_s$, $\mathbb{T}$ is the time-ordering operator and $\tau_p$ is the gate time. One \textit{standard} approach for computing Eq.~(\ref{eqn:OffRes-Def of Ui(t,t0)}) is the Magnus method \cite{Magnus_Exponential_1954, Blanes_Magnus_2009, Blanes_Pedagogical_2010}, which solves perturbatively for the generator of time evolution operator as $\hat{U}_I(\tau_p,0)\equiv \exp[-i \hat{K}(\tau_p,0)]$. Up to the 2nd order one finds   
\begin{subequations}
\begin{align}
& \hat{K}_1(\tau_p,0)=\int_{0}^{\tau_p} dt' \HO_{I}(t')\;,
\label{eqn:OffRes-Def of K1(tau,0)}\\
& \hat{K}_2(\tau_p,0)= -\frac{i}{2} \int_{0}^{\tau_p} dt' \int_{0}^{t'} dt'' [\HO_{I}(t'),\HO_{I}(t'')]\;.
\label{eqn:OffRes-Def of K2(tau,0)}
\end{align}
\end{subequations}
For numerical simulations, we employ a 2nd order Magnus solver \cite{Hairer_Geometric_2006}, based on the discrete form of Eqs.~(\ref{eqn:OffRes-Def of K1(tau,0)})--(\ref{eqn:OffRes-Def of K2(tau,0)}), similar to Ref.~\cite{Tripathi_Operation_2019}. 

The Magnus method computes the two interaction categories, resonant and off-resonant, altogether. Time-dependent SWPT, however, separates the two by computing an effective \textit{resonant} Hamiltonian as
\begin{align}
\hat{\mathcal{H}}_{\text{I,eff}}(t)\equiv \hat{U}_{\text{SW}}^{\dag}(t)\left[\hat{\mathcal{H}}_{\text{I}}(t)-i \partial_t \right]\hat{U}_{\text{SW}}(t) \;,
\label{eqn:OffRes-Def of H_I,eff}
\end{align}
using the frame transformation $\hat{U}_{\text{SW}}(t)\equiv \exp[-i \hat{G}(t)]$. Similar to Magnus, we can solve for the generator $\hat{G}(t)$ and the effective Hamiltonian $\hat{\mathcal{H}}_{\text{I,eff}}(t)$ perturbatively \cite{Boissonneault_Dispersive_2009, Gambetta_Analytic_2011, Magesan_Effective_2020, Malekakhlagh_Lifetime_2020, Petrescu_Lifetime_2020, Malekakhlagh_First-Principles_2020, Petrescu_Accurate_2021} (see Appendices~\ref{App:SWPT} and~\ref{App:EffCRHam}). Therefore, SWPT is a method for implementing \textit{systematic} RWA: through a perturbative frame transformation we obtain effective models with resonant interaction rates, however, the information about off-resonant processes is stored in $\hat{U}_{\text{SW}}(t)$ and hence we can reconstruct the overall time evolution operator as (see Appendices~\ref{App:EffU} and~\ref{App:NonBDU})
\begin{align}
\hat{U}_{\text{I}}(\tau_p,0)=\hat{U}_{\text{SW}}(\tau_p)\hat{U}_{\text{I,eff}}(\tau_p,0)\hat{U}_{\text{SW}}^{\dag}(0) \;.	
\label{eqn:OffRes-UI ITO UIeff}
\end{align}
Equation~(\ref{eqn:OffRes-UI ITO UIeff}) is the bridge between time-dependent Magnus and SWPT formalisms. 

According to Eq.~(\ref{eqn:OffRes-UI ITO UIeff}), the overall time evolution is in principle \textit{invariant} of the frame choice. However, in practice, any perturbative treatments of $\hat{U}_{\text{SW}}(t)$ breaks the invariance. The choice for an efficient frame depends on the drive scheme and the quantities we intend to compute. For CR, to compute an effective Hamiltonian with resonant interactions, the SW frame is BD with respect to the control qubit \cite{Magesan_Effective_2020, Malekakhlagh_First-Principles_2020} so as to capture resonant $X$ and $Y$ target rotations (see Fig.~\ref{fig:OffRes-BDFrame}). Hence, the SW frame transformation $\hat{U}_{\text{SW}}(t)$ encodes the details of off-resonant processes and non-BD interactions in particular.  

Generally, analytical modeling of time-dependent error requires precise and hence very involved symbolic computer algebra as discussed in the Appendices. We base our analysis primarily on the numerical simulation, while using perturbation theory to corroborate various trends of off-resonant error. For instance, the leading order perturbative estimate for the probability amplitude of $\ket{0_c}\rightarrow\ket{1_c}$, with time-dependent coupling $\Omega_c(t)/2$ and detuning $\Delta_{ct}$, over time interval $[0,\tau_p]$ reads (see Appendix E) 
\begin{align}
\begin{split}
&-i \int_{0}^{\tau_p} dt'\frac{\Omega_c(t')}{2} e^{i \Delta_{ct} t'}\\
=&-\int_{-\infty}^{\infty} \frac{d\omega'}{2\pi}\frac{\tilde{\Omega}_c(\omega')}{2(\omega'+\Delta_{ct})}\left[e^{i(\omega'+\Delta_{ct})\tau_p}-1\right] \\
=&- \left. \left\{\sum\limits_{n=0}^{\infty}\left[\frac{1}{\Delta_{ct}}\left(\frac{i}{\Delta_{ct}}\frac{d}{dt'}\right)^n \frac{\Omega_c(t')}{2}\right]e^{i\Delta_{ct} t'}\right\} \right|_{0}^{\tau_p}\;.
\end{split}
\label{eqn:OffRes-1PhProbAmpViaSWPT}
\end{align}        

The first line in Eq.~(\ref{eqn:OffRes-1PhProbAmpViaSWPT}) is an example of a \textit{time-domain} overlap integral between the pulse and transition frequency $\Delta_{ct}$. The second line shows the \textit{frequency-domain} representation, with pulse Fourier transform $\tilde{\Omega}_c(\omega')$, where a single sideband photon provides the energy to excite the transition. From design perspective, this suggests that mitigating the error requires \textit{filtering} (notching) the pulse spectrum at $\omega'=-\Delta_{ct}$. Lastly, assuming that the pulse is differentiable up to arbitrary orders at the boundaries, the third line shows the spectral overlap in terms of the pulse time derivatives through \textit{adiabatic expansion}, and suggests that DRAG \cite{Motzoi_Simple_2009, Gambetta_Analytic_2011} is a natural leading order solution for engineering spectral content (see Appendix~\ref{App:OffResQu}). Similarly, higher order expansions in Magnus and SWPT describe probability amplitude of multi-level off-resonant transitions as processes in which multiple sideband photons provide the total energy for the transition (see the 2nd row in Table~\ref{tab:OffResErrSummary}).

\begin{figure}
\centering
\includegraphics[scale=0.224]{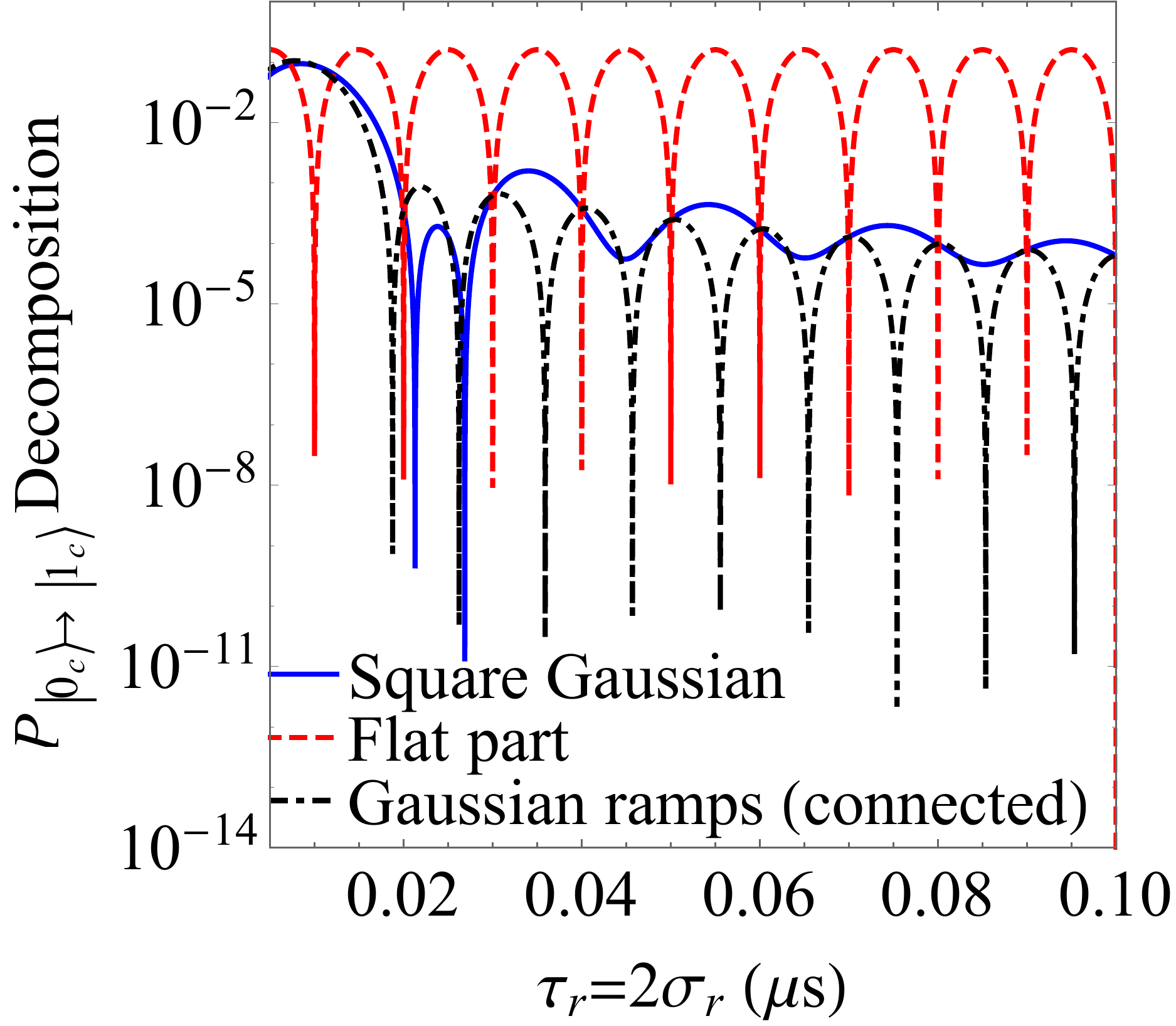}
\includegraphics[scale=0.224]{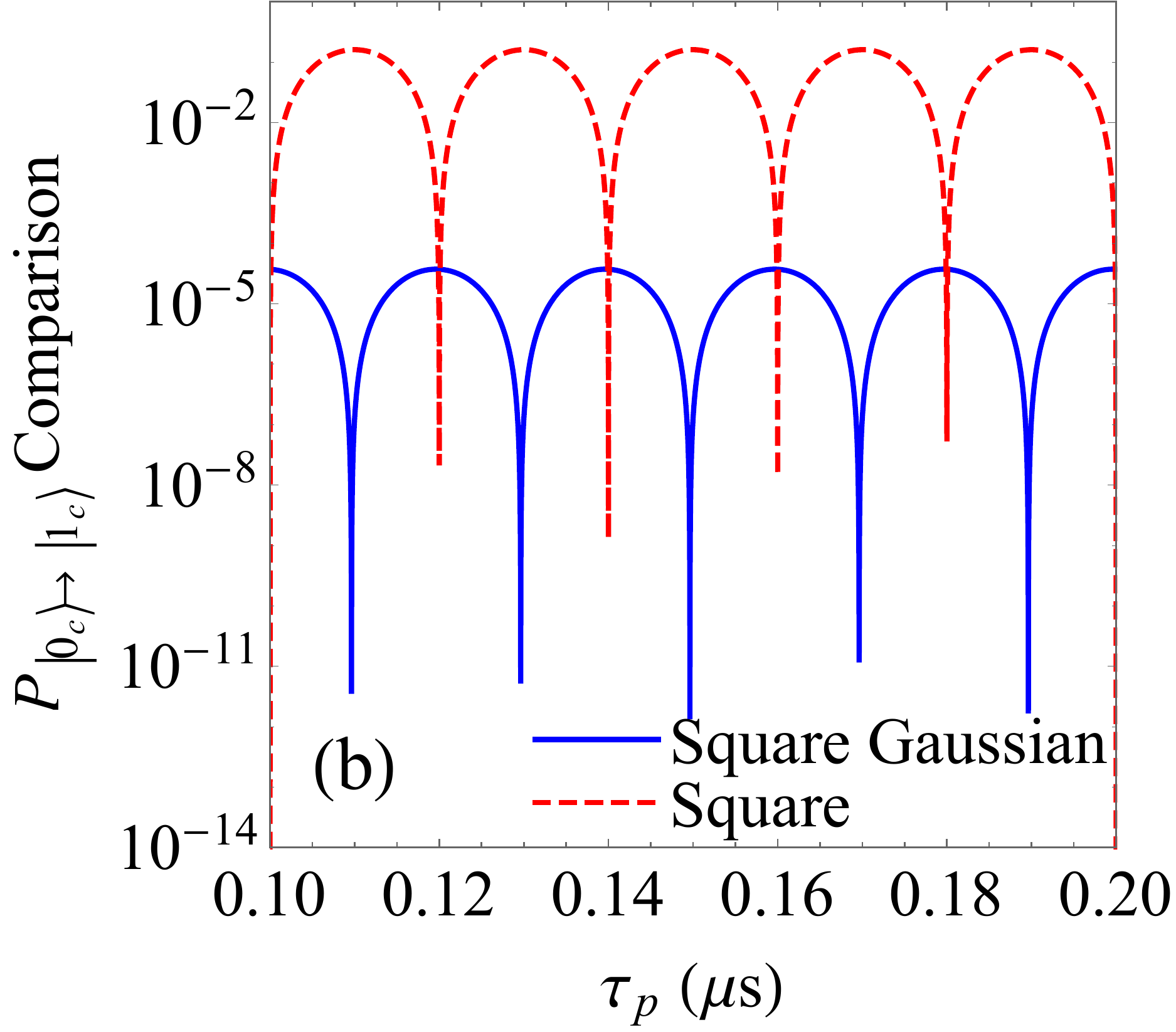}
\caption{(a) Decomposition of the perturbative overlap probability based on Eq.~(\ref{eqn:OffRes-1PhProbAmpViaSWPT}) into overlap with a square Gaussian pulse in Eq.~(\ref{eqn:OffRes-Def of SGPulse}), overlap with just the Gaussian ramps (connected), and overlap with just the flat part as a function of $\tau_r=2\sigma_r$ for fixed $\Omega=20$ MHz, $\Delta_{ct}=50$ MHz and $\tau_p=200$ ns. The distinct overlaps add at the level of complex probability amplitude, hence there is significant interference between separate parts of the pulse. The comparison confirms the crucial role of smooth ramps in reducing the error. (b) Comparison of overlap probability between square Gaussian, for the same parameters as in (a) and fixed $\tau_r=2\sigma_r=26$ ns [an optimal value in panel (a)], and a square pulse with the same $\tau_p$ as a function of $\tau_p$. The results exhibit a periodic dependence on $\tau_p$ with period $2\pi/\Delta_{ct}$.}
\label{fig:OffRes-RampFlatDecomposition}
\end{figure}
It is important to note that off-resonant error is \textit{not} a creature of just the pulse ramps and is present for a constant-amplitude drive. This can be seen from Eq.~(\ref{eqn:OffRes-1PhProbAmpViaSWPT}) where a constant pulse with amplitude $\Omega$ and duration $\tau_p$ results in error probability of $(\Omega/\Delta_{ct})^2\sin^2(\Delta_{ct} \tau_p/2)$, which has a period of $2\pi/\Delta_{ct}$ in $\tau_p$. Smoother ramps, however, bring additional non-trivial derivative contributions that modify and in particular mitigate the error. To see this, we compare to the square Gaussian pulse defined as
\begin{align}
\Omega_{\text{SG}}(t)\equiv
\begin{cases}
\Omega\frac{e^{-\frac{(t-\tau_r)^2}{2\sigma_r^2}}-e^{-\frac{\tau_r^2}{2\sigma_r^2}}}{1-e^{-\frac{\tau_r^2}{2\sigma_r^2}}}  \;, &0<t<\tau_r\\
\Omega\;,  &\tau_r<t<\tau_p-\tau_r\\
\Omega \frac{e^{-\frac{[t-(\tau_p-\tau_r)]^2}{2\sigma_r^2}}-e^{-\frac{\tau_r^2}{2\sigma_r^2}}}{1-e^{-\frac{\tau_r^2}{2\sigma_r^2}}} \;, &\tau_p-\tau_r<t<\tau_p
\end{cases}
\label{eqn:OffRes-Def of SGPulse}
\end{align}
with ramps comprised of a truncated Gaussian with rise time $\tau_r$, standard deviation $\sigma_r$ and a total flat time of $\tau_p-2\tau_r$. Based on the first line of Eq.~(\ref{eqn:OffRes-1PhProbAmpViaSWPT}), contributions from different parts of the pulse, i.e. the ramps and the flat part, add with \textit{complex} amplitude and can create constructive/destructive interference. This is shown in Fig.~\ref{fig:OffRes-RampFlatDecomposition}, where we see that the overlap \textit{probability} with square Gaussian lies almost in between the individual overlap probabilities due to just the ramps or just the flat part. In particular, the overlap due to the flat part is orders of magnitude higher than the overall overlap with square Gaussian pulse, demonstrating the benefit of smooth Gaussian ramps in reducing off-resonant error.   
    
Given that CR gate operates typically in the near-detuned straddling regime ($|\Delta_{ct}|<|\alpha_c|$), the off-resonant error comprises of \textit{three dominant} types involving the following transitions on the control qubit: (1) $\ket{0_c}\rightarrow \ket{1_c}$ with single-photon transition frequency $\Delta_{ct}$, (2) $\ket{0_c}\rightarrow \ket{2_c}$ with two-photon transition frequency $2\Delta_{ct}+\alpha_c$ and (3) $\ket{1_c}\rightarrow \ket{2_c}$ with single-photon transition frequency $\Delta_{ct}+\alpha_c$ (see Table~\ref{tab:OffResErrSummary}). Depending on qubit-qubit detuning, and proximity to the underlying frequency collisions at $\Delta_{ct}=0, \ -\alpha_c/2, \ -\alpha_c$ \cite{Tripathi_Operation_2019, Malekakhlagh_First-Principles_2020}, each error type can become the most dominant. For instance, type 1 (non-BD) error is dominant for qubit pairs with relatively small detuning where $|\Delta_{ct}|\ll |\Delta_{ct}+\alpha_c/2|$ and $|\Delta_{ct}|\ll|\Delta_{ct}+\alpha_c|$.  In the following, we analyze these dominant off-resonant error types numerically.

\subsection{Simulation}
\label{SubSec:OffResSimulation}

To this aim, we employ a 2nd order Magnus solver \cite{Hairer_Geometric_2006} based on the discrete form of Eqs.~(\ref{eqn:OffRes-Def of K1(tau,0)})--(\ref{eqn:OffRes-Def of K2(tau,0)}). We keep 5 and 3 levels for the control and the target qubits, respectively. The main CR pulse, applied on the $X$ axis of the control qubit, is taken to be the square Gaussian in Eq.~(\ref{eqn:OffRes-Def of SGPulse}). Drive amplitudes on the control and the target qubits should then be calibrated according to Eqs.~(\ref{eqn:DirCNOT-Cond IX+ZX=0})--(\ref{eqn:DirCNOT-Cond IX-ZX=Pi}). Here, we use the perturbative conditions~(\ref{eqn:DirCNOT-PertCond IX+ZX=0})--(\ref{eqn:DirCNOT-PertCond IX-ZX=Pi}) to expedite the numerical computation.

\begin{figure}
\centering
\includegraphics[scale=0.1665]{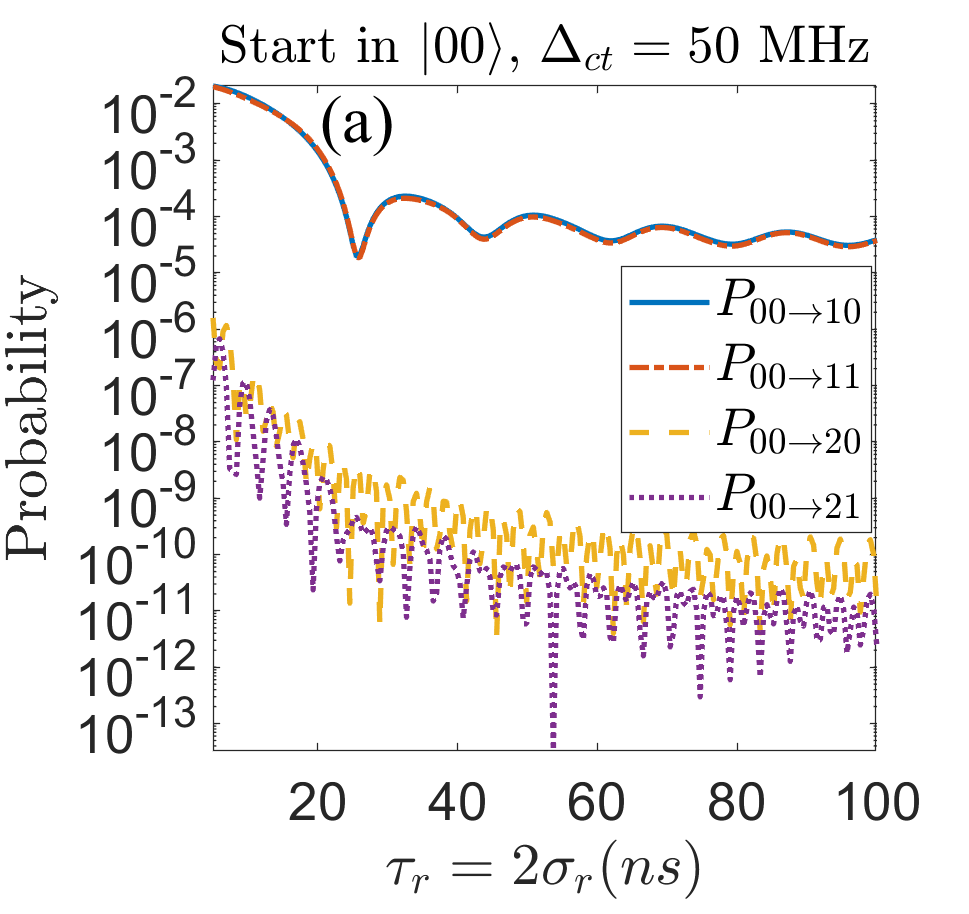}
\includegraphics[scale=0.168]{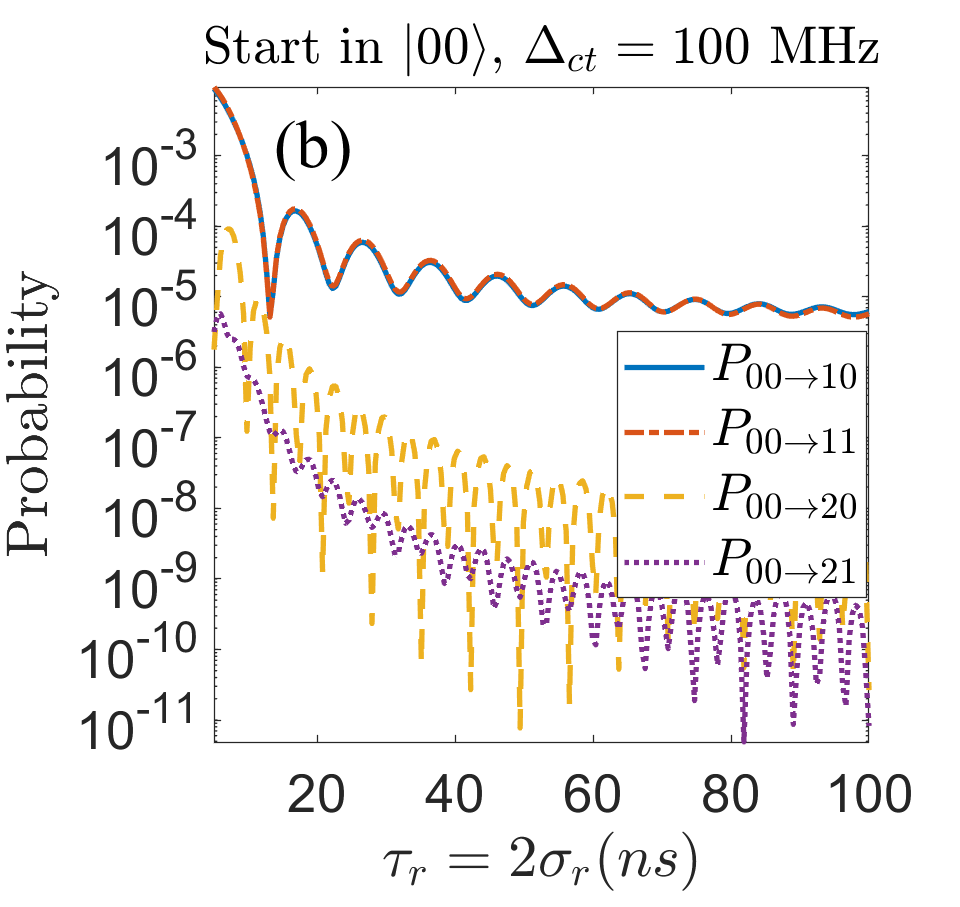}\\
\includegraphics[scale=0.169]{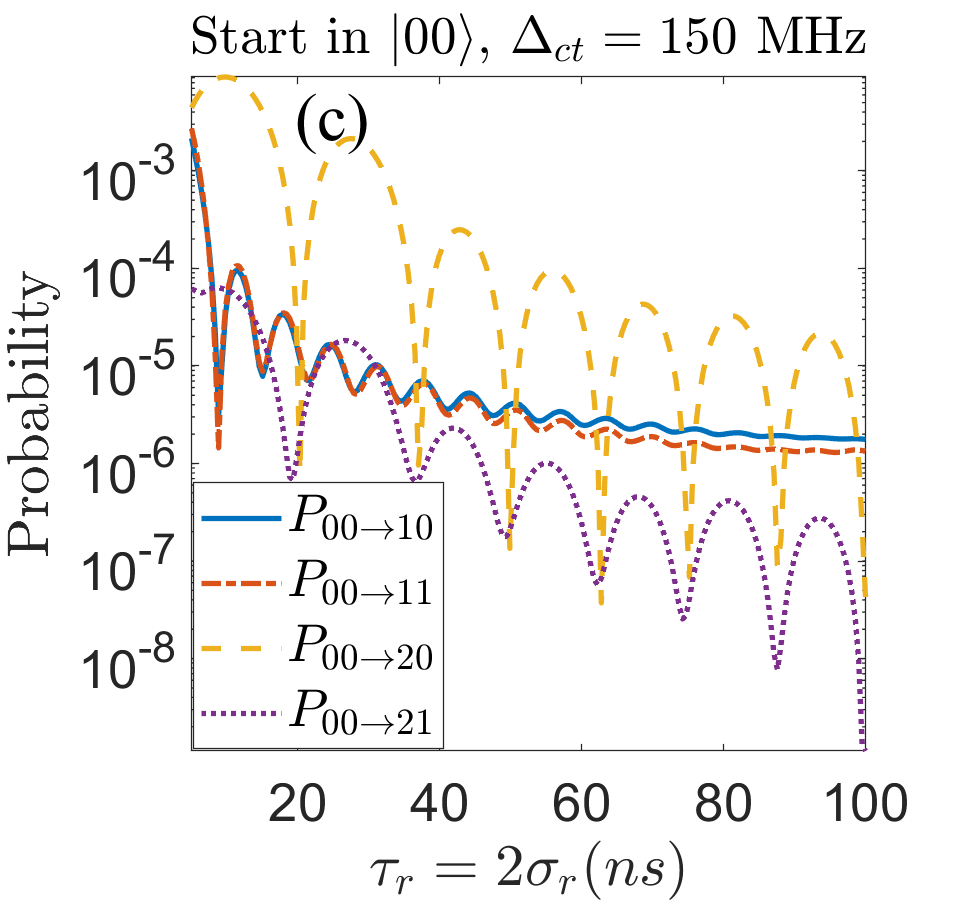}
\includegraphics[scale=0.169]{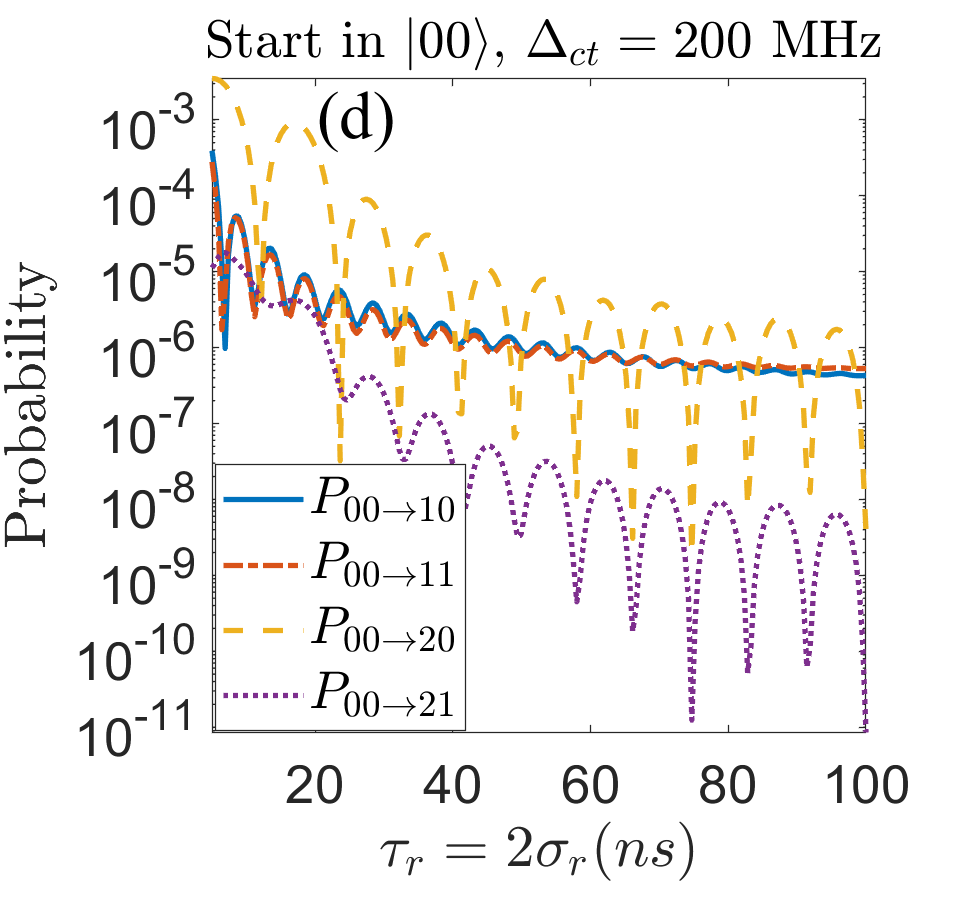}
\includegraphics[scale=0.139]{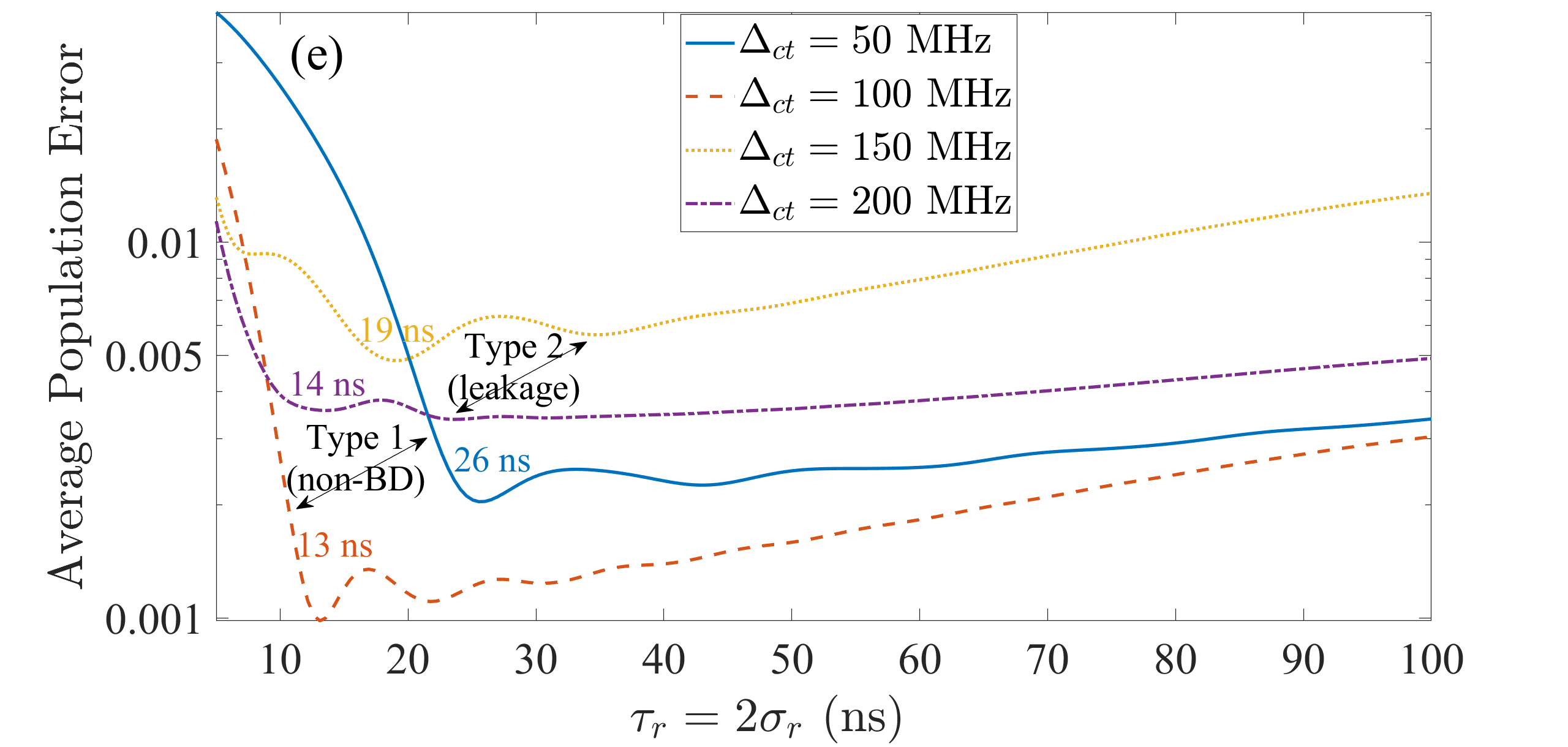}
\caption{Numerical dependence of off-resonant error on $\tau_r=2\sigma_r$ based on the square Gaussian pulse of Eq.~(\ref{eqn:OffRes-Def of SGPulse}) with fixed gate time $\tau_p=200$ ns. System parameters are set as $J=3.5$, $\omega_t=5000$, $\alpha_c=\alpha_t=-340$ MHz, while varying control qubit freuqency for detunings $\Delta_{ct}=50, 100, 150$ and $200$ MHz. The corresponding static $ZZ$ rates are apprximatley $147.2$, $157.7$, $178.8$ and $220.4$  KHz, respectively. (a)--(d) Decomposition of error into individual transitions for initial state set to $\ket{00}$ as a point of comparison. Legends use the shorthand notation $P_{mn\rightarrow pq} \equiv |\bra{pq}\hat{U}_{I}(\tau_p,0)\ket{mn}|^2$. (e) Average population error in Eq.~(\ref{eqn:OffRes-Def of Epop}) calculated using the four computational states.}
\label{fig:OffRes-FuncOfRiseTime}
\end{figure}

Figure~\ref{fig:OffRes-FuncOfRiseTime} shows the behavior of off-resonant error as a function of pulse rise time $\tau_r$ for sample detunings $\Delta_{ct}=50$, $100$, $150$, $200$ MHz, and $\alpha_c=\alpha_t=-340$ MHz. First, depending on the closeness to the three collision types in Sec.~\ref{SubSec:OffResTheory}, we observe a crossover where either non-BD error or leakage become dominant [Figs.~\ref{fig:OffRes-FuncOfRiseTime}(a)--\ref{fig:OffRes-FuncOfRiseTime}(d)]. For instance, for $\Delta_{ct}=50$ MHz, the non-BD error is almost four orders of magnitude larger than leakage, due to proximity to a type 1 collision. Second, there are \textit{favorable} local minima of the off-resonant error, as a function of $\tau_r$, due to overlap of the underlying transition frequency with a dip in the sidebands of the pulse spectrum. In Fig.~\ref{fig:OffRes-FuncOfRiseTime}(e), we show average population error defined as     
\begin{align}
\bar{E}_{\text{pop}}\equiv 1-\frac{1}{4}\sum\limits_{j,k=0,1} \left|\bra{j,k}\hat{U}_{\text{CNOT}}^{\dag}\hat{U}_I(\tau_p,0)\ket{j,k} \right|^2 .
\label{eqn:OffRes-Def of Epop}
\end{align}
We use this measure intentionally to reflect the behavior of off-resonant error more clearly, instead of the overall average error \cite{Pedersen_Fidelity_2007, Magesan_Scalable_2011} that contains phase error due to Stark shifts and $ZZ$. In particular, the average population error follows similar local minima as a function of $\tau_r$ that is dominated by one or interplay of multiple collision types. Based on Fig.~\ref{fig:OffRes-FuncOfRiseTime}(e), possible optimal choices of $\tau_r$ for $\Delta_{ct}=50$, $100$, $150$ and $200$ MHz are \textit{approximately} $26$, $13$, $19$ and $14$ ns, respectively. Moreover, the case of $\Delta_{ct}=100$ MHz leads to the smallest off-resonant error due to being comparably furthest from the three collisions.   

\begin{figure}
\centering
\includegraphics[scale=0.165]{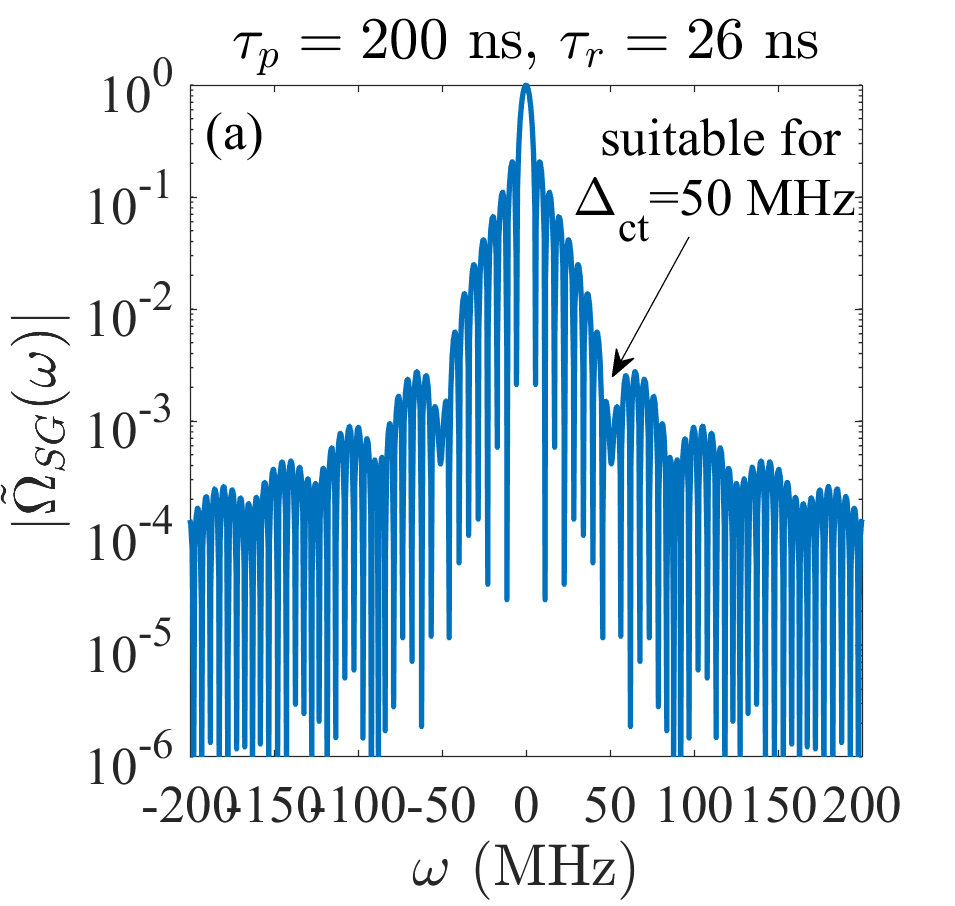}
\includegraphics[scale=0.29]{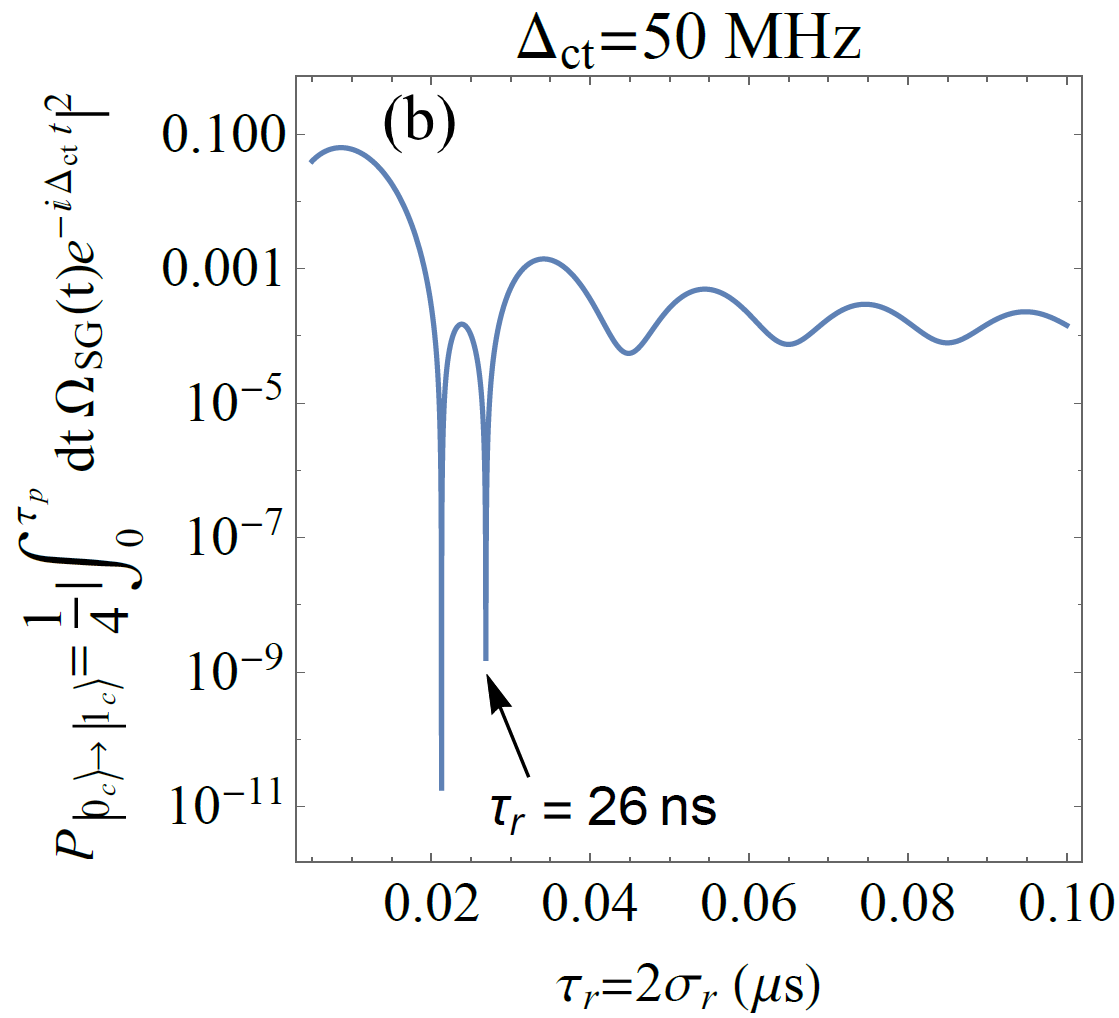}
\caption{(a) Normalized Fourier transform of the square Gaussian pulse~(\ref{eqn:OffRes-Def of SGPulse}) for $\tau_p=200$ ns and $\tau_r=2\sigma_r=26$ ns. (b) Analytical reconstruction of the $\ket{0_c}\rightarrow \ket{1_c}$ error as a function of $\tau_r$ based on the first row of Table~\ref{tab:OffResErrSummary}. Compared to Fig.~\ref{fig:OffRes-RampFlatDecomposition}(a), the drive is adjusted at each $\tau_r$ for an approximate CNOT calibration.}
\label{fig:OffRes-GaussSqFTrans}
\end{figure}

To connect the simulation results to the analytical overlap integrals of Table~\ref{tab:OffResErrSummary}, note that Fourier transform of square Gaussian consists of an overall tail, with relatively \textit{wide} sidebands in frequency, whose spectral width is primarily determined by the risetime $\tau_r$, and a series of \textit{narrower} sidebands with widths determined by interplay between the overall gate time $\tau_p$ and flat time $\tau_p-2\tau_r$ [see Fig.~\ref{fig:OffRes-GaussSqFTrans}(a)]. Increasing $\tau_r$, on the one hand, shrinks the overall spectral width and suppresses the error in general. However, to reach the same CR rotation angle, a stronger drive is needed. Up to the leading order, one- and two-photon off-resonant errors are $O(\Omega_{cx}^2)$ and $O(\Omega_{cx}^4)$, respectively. Therefore, for each parameter set, we expect distinct sweet spots for off-resonant error in terms of $\tau_r$. We took a closer look into the case of $\Delta_{ct}=50$ MHz in Fig.~\ref{fig:OffRes-GaussSqFTrans}. First, Fourier transform of Eq.~(\ref{eqn:OffRes-Def of SGPulse}) for $\tau_p=200$ ns and $\tau_r=26$ ns exhibits a sideband dip at $\omega=50$ MHz justifying why this choice is suitable for suppressing non-BD error in Fig.~\ref{fig:OffRes-FuncOfRiseTime}(a). Moreover, we used the leading order analytical estimate for $\ket{0_c} \rightarrow \ket{1_c}$ in Table~\ref{tab:OffResErrSummary} to reconstruct the dependence on $\tau_r$. We find that the position of local minima agree approximately between the numerical and leading order analytical results in Figs.~\ref{fig:OffRes-FuncOfRiseTime}(a) and~\ref{fig:OffRes-GaussSqFTrans}(b). 

\section{DRAG}
\label{Sec:DRAG}

\begin{figure}[t!]
\includegraphics[scale=0.405]{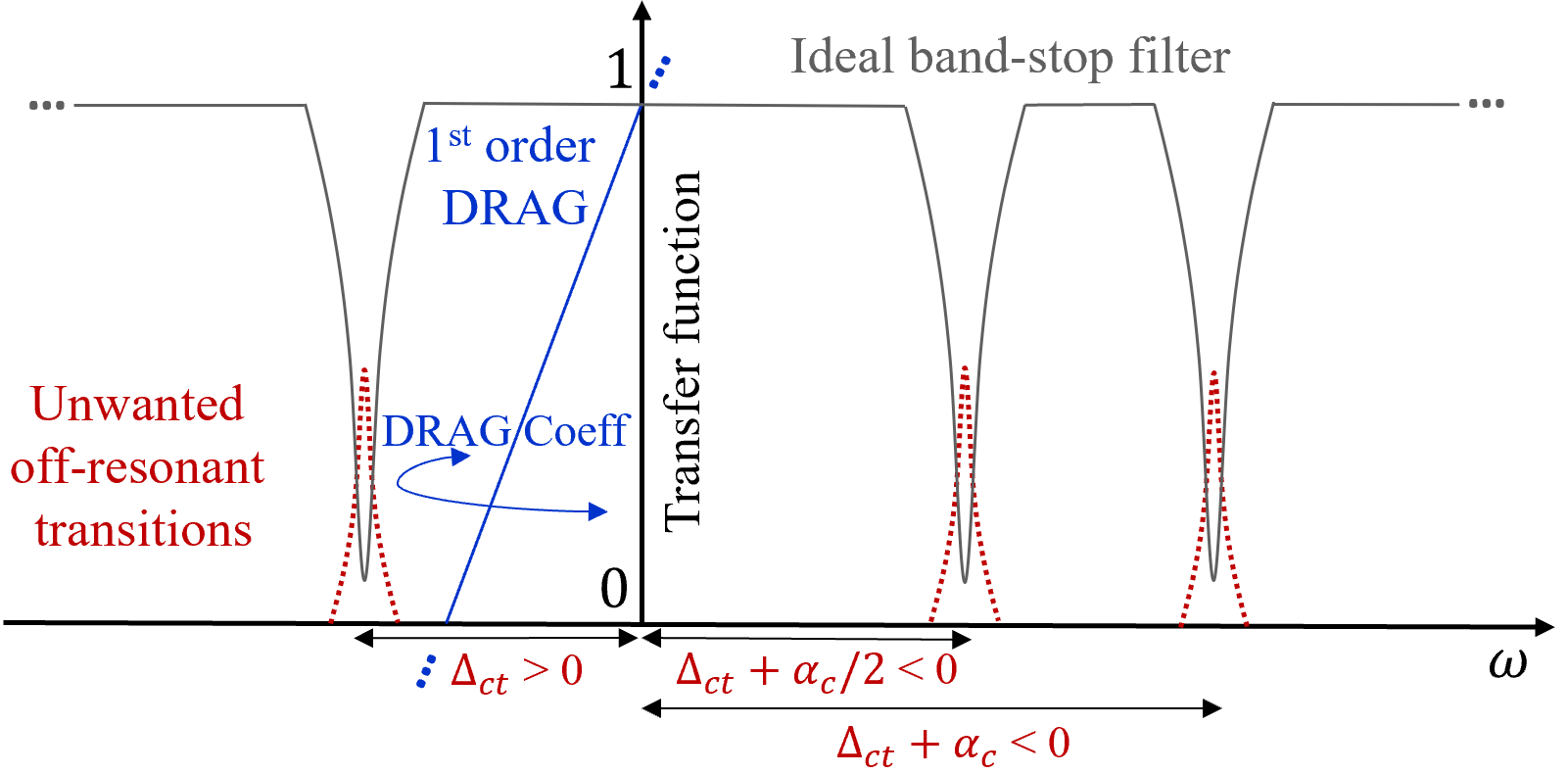}
\centering
\caption{Schematic transfer function of an ideal band-stop filter that mitigates off-resonant error by notching every individual collisions versus a 1st order DRAG solution of the form $\Omega_{cy}(t)=(1/\Delta_D)\dot{\Omega}_{cx}(t)$ and corresponding transfer function $T_{D}(\omega)=1-\omega/\Delta_D$. The frequency allocation here is for a case where non-BD error is most detrimental, i.e. $\Delta_{ct}>0$ is the smallest transition frequency in absolute value. The idea, however, remains the same for other allocations, while optimal DRAG parameter $\Delta_{D}$ varies.}
\label{fig:DRAGvsBSFilterSchematic}
\end{figure}

Local minima of the square Gaussian sidebands act as an \textit{intrinsic} filter for off-resonant transitions as shown in Figs.~\ref{fig:OffRes-FuncOfRiseTime}--\ref{fig:OffRes-GaussSqFTrans}. An ideal frequency composition of a CR pulse should have minimal frequency overlap with the aforementioned three collision types, and possibly higher order transitions for stronger drive. This can in principle be achieved by applying a band-stop filter to the input pulse as     
\begin{align}
\tilde{\Omega}_c(\omega)=T(\omega)\tilde{\Omega}_{\text{SG}}(\omega) \;,
\label{eqn:DRAG-Def of BSFilter}
\end{align}    
where the transfer function $T(\omega)$ should notch every unwanted off-resonant transitions (see Fig.~\ref{fig:DRAGvsBSFilterSchematic}).
It is typically challenging to design a practical band-stop filter in hardware with sufficiently high quality factor.

Here, we explore the application of a DRAG pulse on the control qubit that results in an effective notch filter. In a leading order $Y$-DRAG solution, we augment the square Gaussian, on the $X$ axis of the control qubit, with the derivative on the $Y$ axis as \cite{Motzoi_Simple_2009, Gambetta_Analytic_2011}
\begin{subequations}
\begin{align}
&\Omega_{cx}(t)=\Omega_{\text{SG}}(t) \;,
\label{eqn:DRAG-Def of XComp}\\
&\Omega_{cy}(t)=\frac{1}{\Delta_{D}}\dot{\Omega}_{\text{SG}}(t) \;,
\label{eqn:DRAG-Def of YComp}
\end{align}
\end{subequations}
with $\Delta_D$ as the DRAG parameter. Hence, the overall pulse on the control qubit takes the form $\Omega_c(t)=\Omega_{cx}(t)+i\Omega_{cy}(t)$, and given that $\partial_t \leftrightarrow i\omega$, the effective DRAG transfer function reads 
\begin{align}
T_{D}(\omega)=1-\omega/\Delta_D \;. 
\label{eqn:DRAG-DRAGTrFunc}
\end{align}
Note that the optimal choice for $\Delta_D$ depends on the CR gate frequency allocation. In particular, due to multiple collision possibilities, tuning the DRAG coefficient to mitigate a near-collision scenario can in principle cause the error due to other transitions to grow. Hence, 1st order DRAG is \textit{not} necessarily the most optimal tool for pulse shaping for CR gate, or systems with multiple collisions, as also pointed out in Ref.~\cite{Schutjens_Single-Qubit_2013} for single qubit gates with spectator qubits. This being said, we demonstrate noticeable improvement, especially for a qubit pair close to a type 1 (non-BD) collision.
\begin{figure}
\centering
\includegraphics[scale=0.1691]{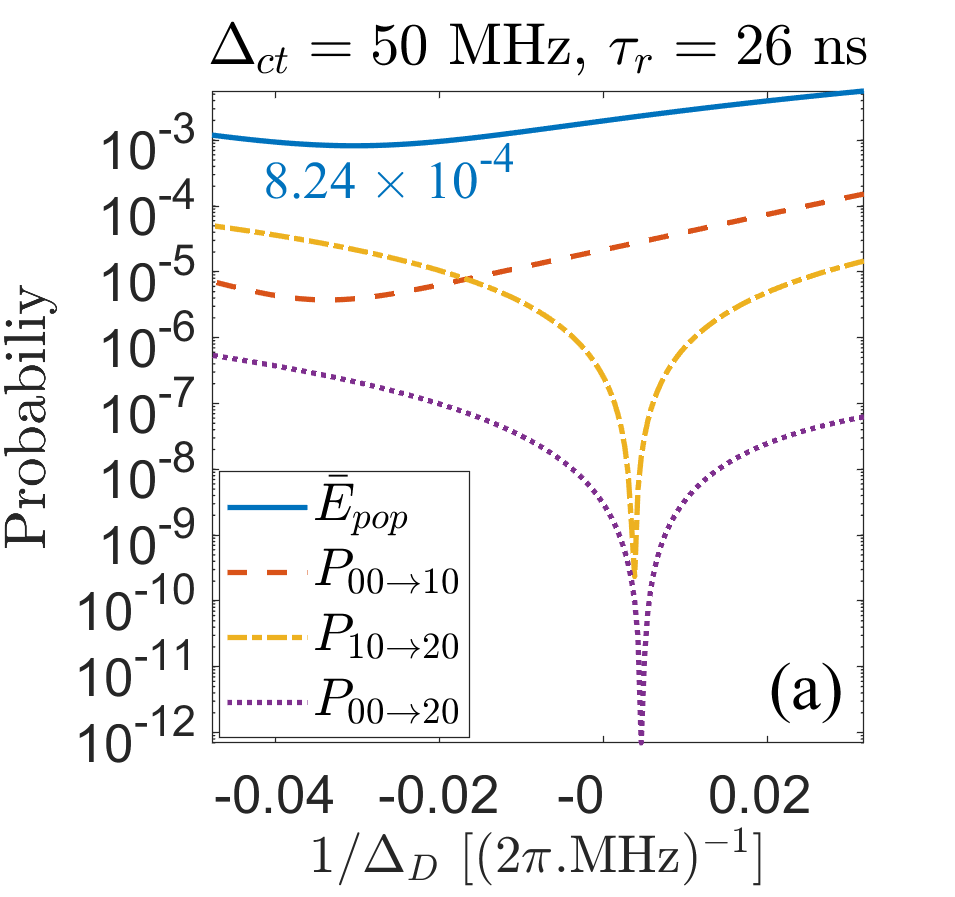}
\includegraphics[scale=0.1691]{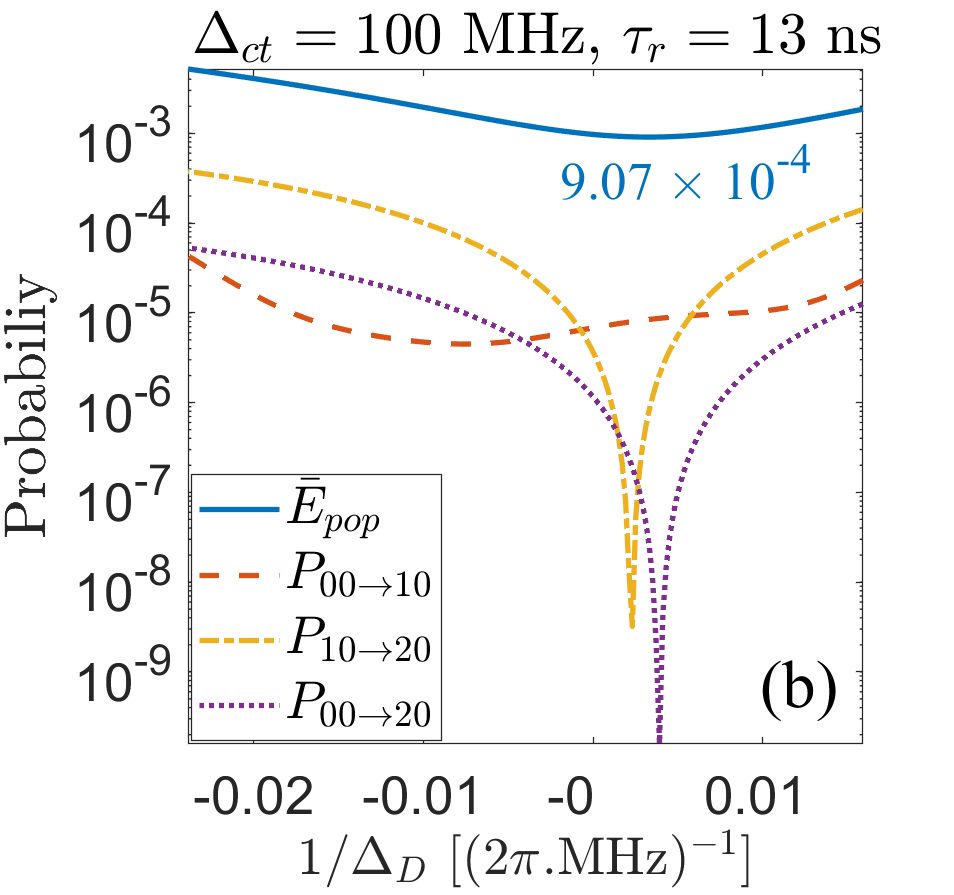}\\
\includegraphics[scale=0.1691]{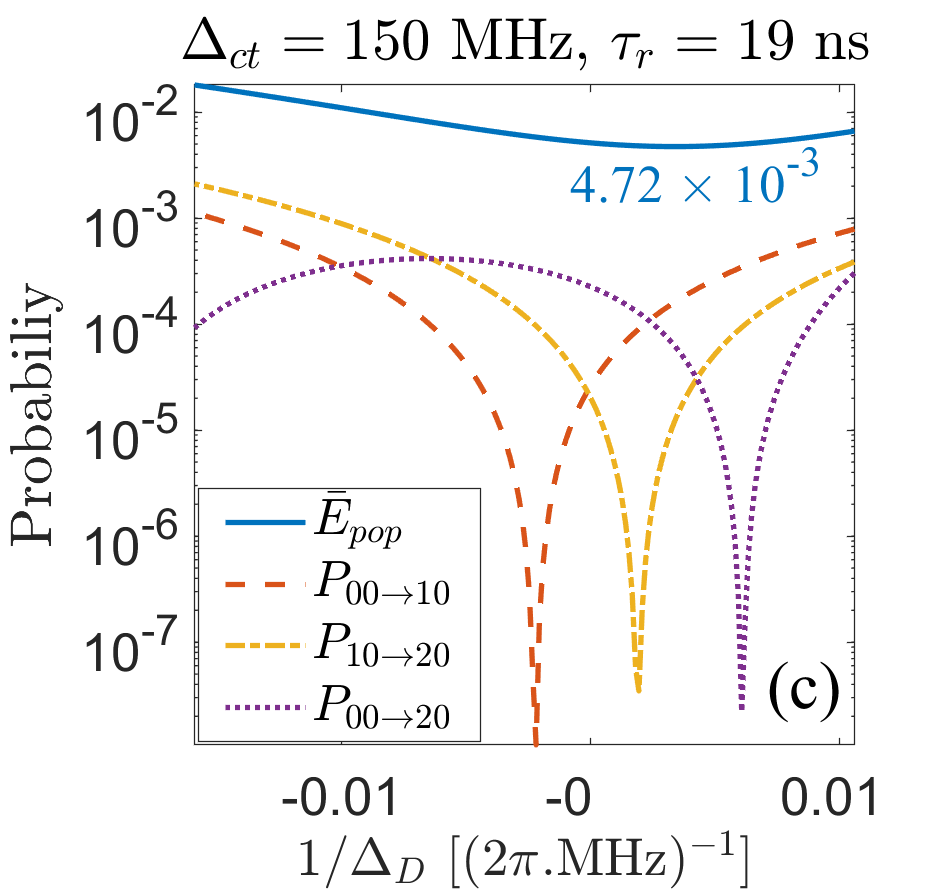}
\includegraphics[scale=0.1691]{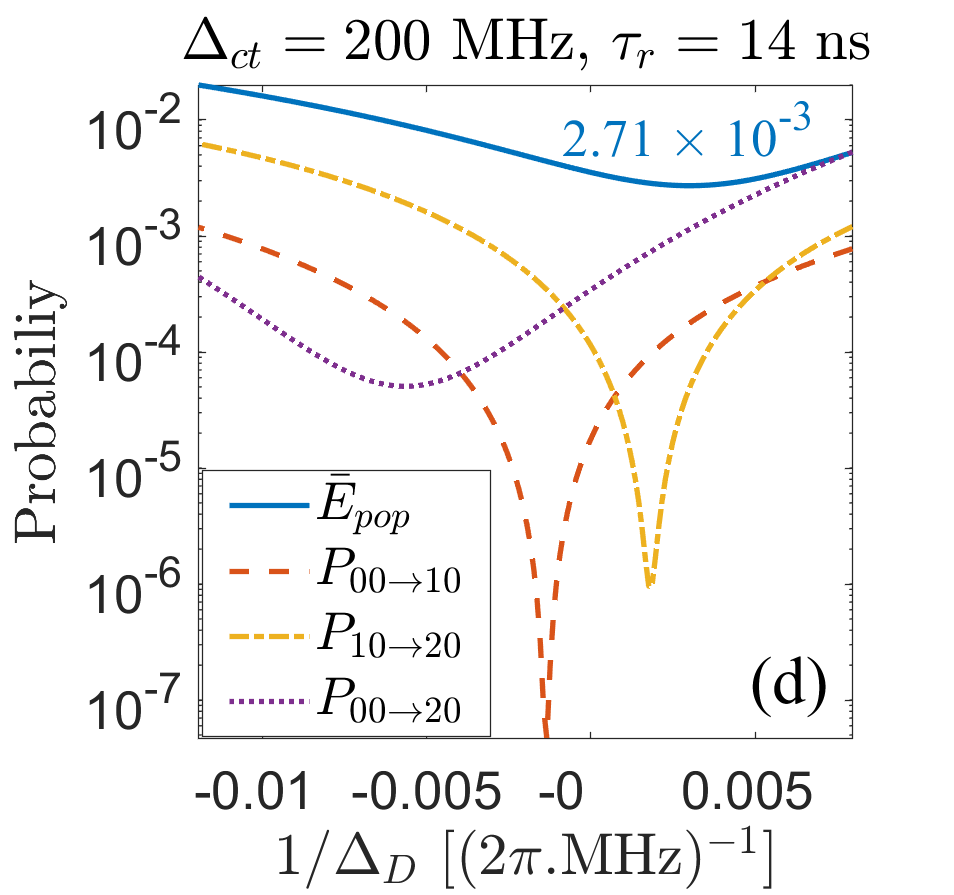}\\
\caption{Distinct off-resonant error types and average population error as a function of DRAG coefficient $1/\Delta_D$ for the same system parameters as in Fig.~\ref{fig:OffRes-FuncOfRiseTime}. The rise time in each case is adopted from the optimal values in Fig.~\ref{fig:OffRes-FuncOfRiseTime}(e) to quantify further improvement by DRAG. The numbers in blue show the average population error at the optimal DRAG parameter for each case.}
\label{fig:DRAGSweep}
\end{figure}

We first discuss how to derive analytical conditions for the DRAG parameter to suppress \textit{individual} collisions in Table~\ref{tab:OffResErrSummary} (see Appendix~\ref{App:OffResQu}). The derivation here is in terms of the leading order term in Magnus, hence slightly distinct from Refs.~\cite{Motzoi_Simple_2009, Gambetta_Analytic_2011}, but reaches similar solutions. For instance, applying adiabatic expansion on the $\ket{0_c}\rightarrow\ket{1_c}$ transition probability gives   
\begin{align}
\begin{split}
&P_{\ket{0_c}\rightarrow\ket{1_c}} \approx \frac{1}{4} \left|\int_{0}^{\tau_p} dt'\Omega_c(t') e^{i \Delta_{ct} t'}\right|^2\\
&=\frac{1}{4} \left|\Big\{\sum\limits_{n=0}^{\infty}\Big[\frac{1}{\Delta_{ct}}\Big(\frac{i}{\Delta_{ct}}\frac{d}{dt'}\Big)^n\Omega_c(t')\Big]e^{i\Delta_{ct} t'}\Big\} \Big|_{0}^{\tau_p}\right|^2 \;.
\end{split}
\label{eqn:DRAG-P0To1 AdiabExp 1}
\end{align}
Replacing the $Y$-DRAG Ansatz~(\ref{eqn:DRAG-Def of XComp})--(\ref{eqn:DRAG-Def of YComp}) into the 2nd line of Eq.~(\ref{eqn:DRAG-P0To1 AdiabExp 1}) results
\begin{align}
\begin{split}
P_{\ket{0_c}\rightarrow\ket{1_c}}=\frac{(1+\lambda_{01})^2}{4\Delta_{ct}^4}\left[\dot{\Omega}_{\text{SG}}^2(\tau_p)+\dot{\Omega}_{\text{SG}}^2(0)\right.
\\
\left.-2\cos(\Delta
_{ct}\tau_p)\dot{\Omega}_{\text{SG}}(\tau_p)\dot{\Omega}_{\text{SG}}(0)\right]+O\left(\ddot{\Omega}_{\text{SG}}^2\right)\;,
\end{split}
\label{eqn:DRAG-P0To1 AdiabExp 2}
\end{align}
with normalized DRAG parameter $\lambda_{01}\equiv \Delta_{ct}/\Delta_{D}$. Higher order terms are given in Appendix~\ref{App:OffResQu}, where we find that the dependence on $\ddot{\Omega}_{\text{SG}}^2$ is also proportional to $(1+\lambda_{01})^2$. Therefore, setting $\lambda_{01}=-1$, i.e. $\Omega_{cy}(t)=-(1/\Delta_{ct})\dot{\Omega}_{\text{SG}}(t)$, removes non-BD error up to terms of $O(\dddot{\Omega}_{\text{SG}}^2)$. Similarly, the leading order DRAG solution for suppressing type 3 error $\ket{1_c}\rightarrow\ket{2_c}$ reads $\Delta_{D}=-(\Delta_{ct}+\alpha_c)$. Two-photon leakage is, however, more involved as the leading order Magnus term appears as a two-time overlap integral (Table~\ref{tab:OffResErrSummary} and Appendix~\ref{App:OffResQu}). Inserting Ansatz~(\ref{eqn:DRAG-Def of XComp})--(\ref{eqn:DRAG-Def of YComp}) results in a 4th order polynomial in $1/\Delta_D$. Intuitively, we expect the optimal DRAG parameter to be approximately set according to \textit{half} of the two-photon transition frequency, i.e. $\Delta_D \approx -(\Delta_{ct}+\alpha_c/2)$. 

\begin{figure}
\centering
\includegraphics[scale=0.171]{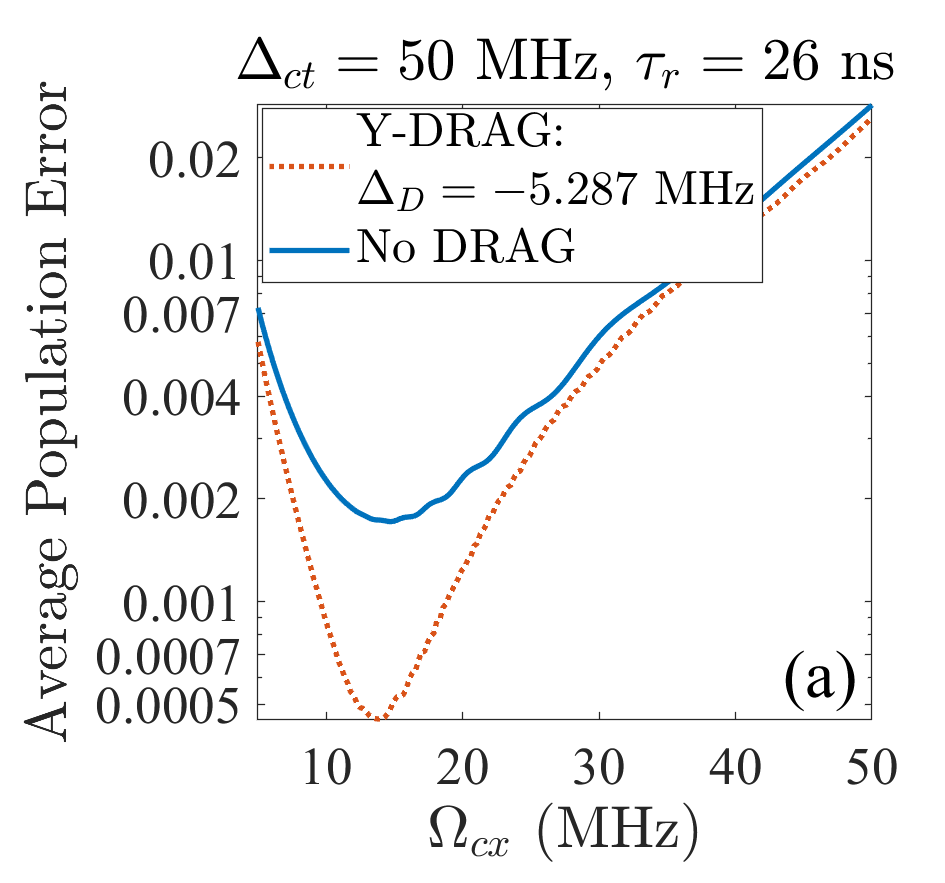}
\includegraphics[scale=0.171]{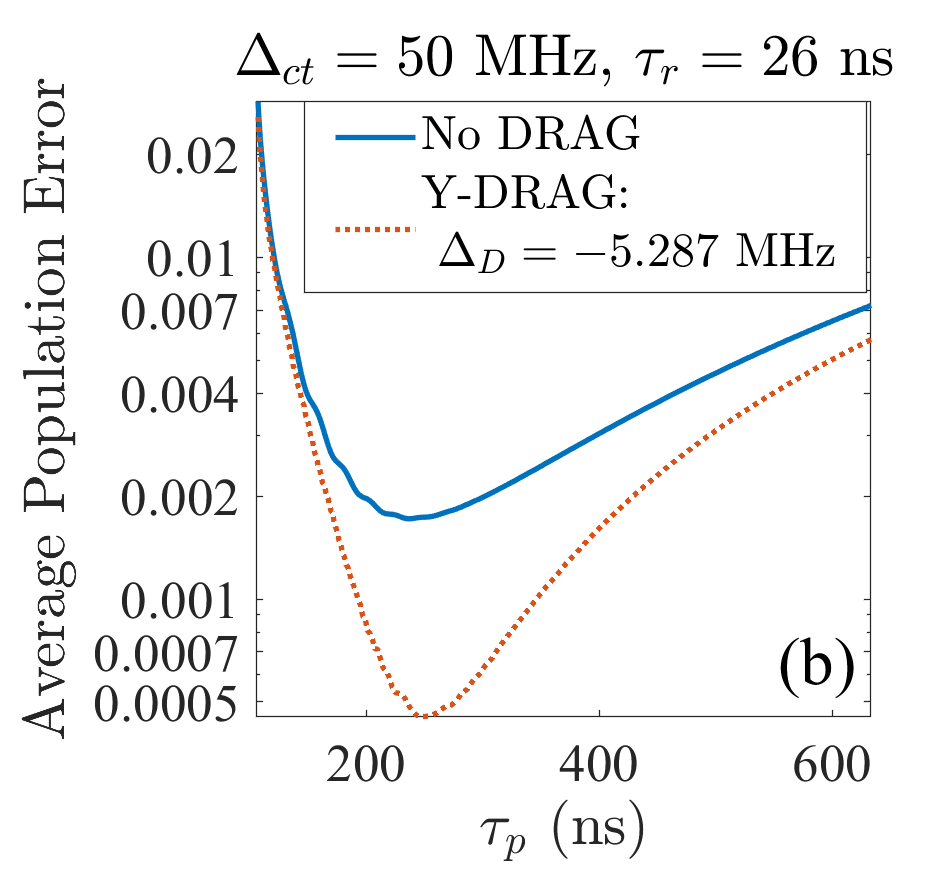}
\caption{Average population error as a function of (a) drive amplitude, and (b) gate time, for the $\Delta_{ct}=50$ MHz case and DRAG parameter set from the optimal value in Fig.~\ref{fig:DRAGSweep}(a) as $\Delta_{D}=-5.287$ MHz. The DRAG correction results in an approximate improvement of average error from $1.72\times 10^{-3}$ down to $4.53\times 10^{-4}$.}
\label{fig:DriveSweepWithWithoutDRAG}
\end{figure}

The above analytical DRAG solutions are based on an off-resonantly driven but isolated transmon qubit, while there are in principle $O(J\Omega_t)$ corrections. Moreover, DRAG correcting for one error type may increase the other (see Fig.~\ref{fig:DRAGvsBSFilterSchematic}). Therefore, we resort to a numerical sweep of the DRAG parameter in order to minimize the \textit{average} population error. Figure~\ref{fig:DRAGSweep} shows $\bar{E}_{\text{pop}}$ along with the three off-resonant error types as a function of $1/\Delta_D$, with $\tau_r$ set to the optimal value based on Fig.~\ref{fig:OffRes-FuncOfRiseTime}(e). We see that depending on $\Delta_{ct}$, there is a trade off between optimal $\Delta_D$ for individual error types as expected from analytics. For instance, the 50 MHz detuned pair in Fig.~\ref{fig:DRAGSweep}(a) is mainly limited by non-BD error without DRAG. Adding DRAG is beneficial in suppressing non-BD error, but increases the leakage to state $\ket{2_c}$. The optimal DRAG parameter is then determined by a balance between the two mechanisms. Lastly, sweeping drive amplitude (gate time) in Fig.~\ref{fig:DriveSweepWithWithoutDRAG} shows partial improvement of the average error with a slower gate of 246 ns. Altogether, with Gaussian shaping and Y-DRAG, the average population error for case (a) can be suppressed down to $4.53\times 10^{-4}$. For pairs close to the type 2 collision, however, the interplay between different transitions is more involved and there is less improvement.     	         

\section{Conclusion}
\label{Sec:Conclusion}

Current calibrations of a CNOT gate using CR architecture are premised on a BD form for effective interactions. In this work, we characterized off-resonant CR interactions, specifically non-BD contributions, as a potential source of coherent error, and illustrated its interplay with control pulse shapes. Time-dependent SW and Magnus perturbations reveal that off-resonant error occurs due to spectral overlap between the pulse and three unwanted off-resonant transitions on the control qubit, denoted as type 1--type 3. Non-BD error is enhanced for pairs in proximity of type 1, while leakage is increased close to type 2--type 3 transitions. 

To suppress such error terms, the pulse spectrum should have minimal content at the underlying transitions. The most immediate solution lies in optimal frequency allocation, i.e. simultaneously maximizing all off-resonant transition frequencies. This requires very precise fabrication of fixed-frequency transmons, where laser annealing \cite{Hertzberg_Laser_2020, Zhang_High_2020} has shown promising improvement. As a more active measure, optimal control techniques can effectively reduce the collision bounds, and allow working with CR pairs in closer proximity to the unwanted transitions. To this aim, we demonstrated promising improvement using two complimentary methods of ramp optimization and Y-DRAG on the control qubit, using a square Gaussian pulse shape. These methods come with little additional cost, i.e. small enhancement of leakage, and are preferable especially since they do not increase the total gate time. More involved optimal pulse shaping is a subject of future research. Our initial experiments on IBM CR processors confirm the benefit of pulse ramp optimization and DRAG on the average error obtained from two-qubit randomized benchmarking \cite{Magesan_Scalable_2011, Magesan_Efficient_2012}. The experimental results will be presented in a subsequent work.          
               
\section{Acknowledgement}
\label{Sec:Acknowledgement}
We appreciate insightful discussions with Emily Pritchett, Ken X Wei, Isaac Lauer, Abhinav Kandala, David C McKay, Daniel Puzzuoli, Oliver E Dial, Seth T Merkel, Maika Takita, Antonio Corcoles and Jay M Gambetta. This work was supported by the Intelligence Advanced Research Projects Activity (IARPA) under contract W911NF-16-1-0114.
 

\appendix
\section{Time-dependent SWPT}
\label{App:SWPT}

Here, we review the main results of a generalized \textit{time-dependent} SWPT formalism~\cite{Malekakhlagh_First-Principles_2020}. We demonstrate the connection between time-dependent and independent perturbation through adiabatic expansion, in which effective rates depend on both the input pulse shapes as well as higher order time derivatives. Furthermore, we discuss how time-dependent SWPT and Magnus expansion are related through a frame transformation. 

Consider a driven quantum system with time-dependent Hamiltonian
\begin{align}
\HO_{s}(t)=\HO_{0}+\lambda\HO_{\text{int}}(t) \;,
\label{Eq:SWPT-Def of lambda}
\end{align}
where $\HO_0$ is the zeroth-order system Hamiltonian, $\HO_\text{int}(t)$ denotes the time-dependent perturbation and $\lambda$ is an auxiliary expansion parameter that facilitates the bookkeeping of perturbative corrections. Moving to the interaction frame with respect to $\HO_0$ we find  
\begin{align}
\lambda\HO_{\text{I}}(t)\equiv e^{i\HO_0 t} \left[\lambda \HO_{\text{int}}(t)\right]e^{-i\HO_0 t} \;,
\label{Eq:SWPT-Def of H_I(t)}
\end{align}
which simplifies the perturbation theory in what follows.

The idea behind SWPT is to average out high-frequency off-resonant processes to come up with effective \textit{resonant} interactions. Formally, this is equivalent to applying a unitary SW transformation to Eq.~(\ref{Eq:SWPT-Def of H_I(t)}) as
\begin{align}
\hat{\mathcal{H}}_{\text{I,eff}}(t)\equiv \hat{U}_{\text{SW}}^{\dag}(t)\left[\lambda \hat{\mathcal{H}}_{\text{I}}(t)-i \partial_t \right]\hat{U}_{\text{SW}}(t) \;,
\label{Eq:SWPT-Def of H_I,eff}
\end{align}
where $\hat{U}_{\text{SW}}(t)\equiv \exp[-i\hat{G}(t)]$, $\hat{G}(t)$ is the unknown SW generator and $\HO_{\text{I,eff}}(t)$ is the effective Hamiltonian of interest. To obtain perturbative solutions, we first perform a series expansion of $\hat{G}(t)$ and $\HO_{\text{I,eff}}(t)$ in terms of $\lambda$ as
\begin{subequations}
\begin{align}
&\hat{G}(t)=\sum\limits_{\lambda=1}^{\infty}\lambda^n \hat{G}_n(t) \;,
\label{Eq:SWPT-G exp}\\
&\HO_{\text{I,eff}}(t)=\sum\limits_{\lambda=1}^{\infty}\lambda^n \HO_{\text{I,eff}}^{(n)}(t) \;.
\label{Eq:SWPT-H_I,eff exp}	
\end{align}
\end{subequations}
Collecting equal powers of $\lambda$ and enforcing the frame change at any arbitrary order, we find \cite{Malekakhlagh_First-Principles_2020}
\begin{subequations}
\begin{align}
O(\lambda):
\begin{cases}
&\hat{\mathcal{H}}_{\text{I,eff}}^{(1)}=\mathcal{B}\left(\hat{\mathcal{H}}_I\right) \;,\\
&\dot{\hat{G}}_1=\mathcal{N}\left(\hat{\mathcal{H}}_I\right) \;,
\end{cases}
\label{Eq:SWPT-O(lambda) HIeff}
\end{align}
\begin{align}
O(\lambda^2):
\begin{cases}
&\hat{\mathcal{H}}_{\text{I,eff}}^{(2)}=\mathcal{B}\left(i[\hat{G}_1,\hat{\mathcal{H}}_I]-\frac{i}{2}[\hat{G}_1,\dot{\hat{G}}_1]\right) \;,\\
&\dot{\hat{G}}_2=\mathcal{N} \left(i[\hat{G}_1,\hat{\mathcal{H}}_I]-\frac{i}{2}[\hat{G}_1,\dot{\hat{G}}_1]\right) \;,
\end{cases}
\label{Eq:SWPT-O(lambda^2) HIeff}
\end{align}
\begin{align}
O(\lambda^3):
\begin{cases}
\begin{split}
\hat{\mathcal{H}}_{\text{I,eff}}^{(3)}&=\mathcal{B}\Big(-\frac{i}{2}[\hat{G}_1,\dot{\hat{G}}_2]-\frac{i}{2}[\hat{G}_2,\dot{\hat{G}}_1]\\
&+\frac{1}{6}[\hat{G}_1,[\hat{G}_1,\dot{\hat{G}}_1]]+i[\hat{G}_2,\hat{\mathcal{H}}_I]\\
&-\frac{1}{2}[\hat{G}_1,[\hat{G}_1,\hat{\mathcal{H}}_I]] \Big)\;,
\end{split}\\
\begin{split}
\dot{\hat{G}}_3&=\mathcal{N}\Big(-\frac{i}{2}[\hat{G}_1,\dot{\hat{G}}_2]-\frac{i}{2}[\hat{G}_2,\dot{\hat{G}}_1]\\
&+\frac{1}{6}[\hat{G}_1,[\hat{G}_1,\dot{\hat{G}}_1]]+i[\hat{G}_2,\hat{\mathcal{H}}_I]\\
&-\frac{1}{2}[\hat{G}_1,[\hat{G}_1,\hat{\mathcal{H}}_I]] \Big)\;,
\end{split}
\end{cases}
\label{Eq:SWPT-O(lambda^3) HIeff}
\end{align}
\begin{align}
O(\lambda^4):
\begin{cases}
\begin{split}
& \hat{\mathcal{H}}_{\text{I,eff}}^{(4)}=\mathcal{B}\Big(-\frac{i}{2}[\hat{G}_1,\dot{\hat{G}}_3]-\frac{i}{2}[\hat{G}_2,\dot{\hat{G}}_2]\\
&-\frac{i}{2}[\hat{G}_3,\dot{\hat{G}}_1]+\frac{1}{6}[\hat{G}_1,[\hat{G}_1,\dot{\hat{G}}_2]]\\
&+\frac{1}{6}[\hat{G}_1,[\hat{G}_2,\dot{\hat{G}}_1]]+\frac{1}{6}[\hat{G}_2,[\hat{G}_1,\dot{\hat{G}}_1]]\\
&+\frac{i}{24}[\hat{G}_1,[\hat{G}_1,[\hat{G}_1,\dot{\hat{G}}_1]]+i[\hat{G}_3,\hat{\mathcal{H}}_I]\\
&-\frac{1}{2}[\hat{G}_1,[\hat{G}_2,\hat{\mathcal{H}}_I]]-\frac{1}{2}[\hat{G}_2,[\hat{G}_1,\hat{\mathcal{H}}_I]]\\
&-\frac{i}{6}[\hat{G}_1,[\hat{G}_1,[\hat{G}_1,\hat{\mathcal{H}}_I]]] \Big)\;,
\end{split}\\
\begin{split}
& \dot{\hat{G}}_4=\mathcal{N}\Big(-\frac{i}{2}[\hat{G}_1,\dot{\hat{G}}_3]-\frac{i}{2}[\hat{G}_2,\dot{\hat{G}}_2]\\
&-\frac{i}{2}[\hat{G}_3,\dot{\hat{G}}_1]+\frac{1}{6}[\hat{G}_1,[\hat{G}_1,\dot{\hat{G}}_2]]\\
&+\frac{1}{6}[\hat{G}_1,[\hat{G}_2,\dot{\hat{G}}_1]]+\frac{1}{6}[\hat{G}_2,[\hat{G}_1,\dot{\hat{G}}_1]]\\
&+\frac{i}{24}[\hat{G}_1,[\hat{G}_1,[\hat{G}_1,\dot{\hat{G}}_1]]+i[\hat{G}_3,\hat{\mathcal{H}}_I]\\
&-\frac{1}{2}[\hat{G}_1,[\hat{G}_2,\hat{\mathcal{H}}_I]]-\frac{1}{2}[\hat{G}_2,[\hat{G}_1,\hat{\mathcal{H}}_I]]\\
&-\frac{i}{6}[\hat{G}_1,[\hat{G}_1,[\hat{G}_1,\hat{\mathcal{H}}_I]]] \Big)\;.
\end{split}
\end{cases}
\label{Eq:SWPT-O(lambda^4) HIeff}
\end{align}
\end{subequations}

In Eqs.~(\ref{Eq:SWPT-O(lambda) HIeff})--(\ref{Eq:SWPT-O(lambda^4) HIeff}), $\mathcal{B} (\bullet)$ and $\mathcal{N}(\bullet)$ represent projections onto the desired frame, in which effective interactions are \textit{resonant}, and the rest of the Hilbert space, respectively. For a CW drive, this corresponds to separating the zero-frequency part of an arbitrary operator $\hat{O}(t)$ [right-hand side of Eqs.~(\ref{Eq:SWPT-O(lambda) HIeff})--(\ref{Eq:SWPT-O(lambda^4) HIeff})] as \cite{Petrescu_Accurate_2021}   
\begin{subequations}
\begin{align}
& \mathcal{B}\left(\hat{O}(t)\right)\equiv \lim_{T \rightarrow \infty}\frac{1}{T} \int_{0}^{T} dt' \hat{O}(t') \;, 
\label{Eq:SWPT-Def of B(.)}\\
&\mathcal{N}\left(\hat{O}(t)\right)\equiv \hat{O}(t)-\mathcal{B}\left(\hat{O}(t)\right)\;.
\label{Eq:SWPT-Def of N(.)}
\end{align}
\end{subequations}
For our CR model in Eqs.~(\ref{eqn:DirCNOT-Def of H0})--(\ref{eqn:DirCNOT-Def of Hd(t)}), and under a CW drive, $\mathcal{B}\left(\bullet\right)$ in Eq.~(\ref{Eq:SWPT-Def of B(.)}) is equivalent to keeping contributions in the BD frame shown in Fig.~\ref{fig:OffRes-BDFrame}. To generalize for a pulsed drive, we adopt the same definition of BD frame as the CW case. The time-dependence due to the pulse shape then appears as overlap integrals between system transition frequencies and the pulse shape. The overlap integrals can also be related to the time-derivatives of the pulse through an adiabatic expansion as discussed in the following.         
  
Note that in solving the operator-valued ODEs~(\ref{Eq:SWPT-O(lambda) HIeff})--(\ref{Eq:SWPT-O(lambda^4) HIeff}), the initial conditions for $\hat{G}_n(t)$ appear as free parameters, where different choices here correspond to distinct frames. The goal of SWPT is to reach a desired form for the effective Hamiltonian by removing off-resonant interactions. Therefore, the \textit{homogeneous} solution is \textit{not} of interest and the natural choice is to solve for the \textit{particular} solution, obtained by the \textit{indefinite} integral of the right hand side at each order. The indefinite integral is consistent with the fact that in SWPT we are interested in effective Hamiltonian rates, while effective rotation angles can be obtained after this step by \textit{definite} integral over pulse duration. To explain this point, we show the explicit expression for the effective Hamiltonian up to the second order. According to Eq.~(\ref{Eq:SWPT-O(lambda) HIeff}), the lowest order generator should be set as
\begin{align}
\hat{G}_1(t)=\int^{t}dt' \mathcal{N}\left(\HO_I(t')\right) \;.
\label{Eq:SWPT-G1 ITO HI}
\end{align}
Replacing solution~(\ref{Eq:SWPT-G1 ITO HI}) into the second order expression~(\ref{Eq:SWPT-O(lambda^2) HIeff}) results 
\begin{align}
\begin{split}
\HO_{\text{I,eff}}^{(2)}(t)&=\mathcal{B}\left(i\left[\int^{t}dt'\mathcal{N}\left(\HO_I(t')\right), \HO_I(t)\right]\right) \\
&-\mathcal{B}\left(\frac{i}{2}\left[\int^{t}dt'\mathcal{N}\left(\HO_I(t')\right), \mathcal{N}\left(\HO_I(t)\right)\right]\right) \;,
\end{split}
\label{Eq:SWPT-HIeff^(2) ITO HI}
\end{align}
which can be understood as possible mixings between interaction Hamiltonian at time $t$ and $t'$. Higher-order solutions account for more complex time correlations. In the limiting case where $\mathcal{B}(\HO_{I})=0$, and hence $\mathcal{N}(\HO_{I})=\HO_{I}$, Eq.~(\ref{Eq:SWPT-HIeff^(2) ITO HI}) simplifies to $\HO_{\text{I,eff}}^{(2)}(t)=\mathcal{B}(\frac{i}{2}[\int^{t}dt'\HO_I(t'), \HO_I(t)])$ as in Ref.~\cite{Malekakhlagh_First-Principles_2020}.	

It is importat to note that, for any driven quantum system, the dynamics is determined by the \textit{overall} time evolution operator and hence is \textit{independent} of our frame choice. Therefore, we need to map the effective unitary operator back to the interaction frame. This step is commonly neglected in SWPT, but is crucial in capturing the error of our effective model due to \textit{off-resonant} processes. Given that the interaction and the effective frame wavefunctions are related as $\ket{\Psi_\text{I}(t)}=\hat{U}_{\text{SW}}(t) \ket{\Psi_\text{I,eff}(t)}$ we find    
\begin{align}
\hat{U}_{\text{I}}(t,t_0)=\hat{U}_{\text{SW}}(t)\hat{U}_{\text{I,eff}}(t,t_0)\hat{U}_{\text{SW}}^{\dag}(t_0) \;,
\label{Eq:SWPT-U_I(t,t0) ITO U_I,eff(t,t0)}
\end{align}
where $\hat{U}_{\text{I,eff}}(t,t_0)$ is given in terms of the effective Hamiltonian $\HO_{\text{I,eff}}(t)$ as
\begin{align}
\hat{U}_{\text{I,eff}}(t,t_0) = \mathbb{T} \exp \left[-i \int_{t_0} ^{t} dt' \HO_{\text{I,eff}}(t')\right] \;.
\label{Eq:SWPT-Def of U_I,eff(t,t0)}
\end{align}
Furthermore, the SW transformation in Eq.~(\ref{Eq:SWPT-U_I(t,t0) ITO U_I,eff(t,t0)}) can be computed by either a matrix exponentiation of the perturbative solution for $\hat{G}$, which is norm-preserving, or by a perturbative expansion of the exponential as  
\begin{align}
\begin{split}
\hat{U}_\text{SW} & = \hat{I}-i\hat{G}_1-i\hat{G}_2-\frac{1}{2}\hat{G}_1^2\\
& -i\hat{G}_3-\frac{1}{2}\left(\hat{G}_1\hat{G}_2+\hat{G}_2\hat{G}_1\right)+\frac{i}{6}\hat{G}_1^3 \\
&-i\hat{G}_4-\frac{1}{2}(\hat{G}_1\hat{G}_3+\hat{G}_3\hat{G}_1) -\hat{G}_2^2 \\
& +\frac{i}{6} \left( \hat{G}_1^2\hat{G}_2 + \hat{G}_1\hat{G}_2\hat{G}_1 + \hat{G}_2\hat{G}_1^2\right)\\
&+\frac{1}{24}\hat{G}_1^4 + O (\lambda^5) \;.
\end{split}
\label{Eq:SWPT-U_SW exp}
\end{align}

Time-dependent SWPT provides the means to also quantify adiabaticity, and importantly, in the limit of adiabatic response, the results agree with those found from time-independent perturbation. In particular, consider a \textit{generic} correction $\int^{t}dt' \Omega(t') \exp[i\Delta(t-t')]$, which is a form that derives from Eq.~(\ref{Eq:SWPT-HIeff^(2) ITO HI}). Here, $\Omega(t)$ denotes the time-dependent drive amplitude with an intrinsic rise time $\tau_r$, and $\Delta$ is the transition frequency for the underlying physical process. Adiabatic response is ensured when the transition frequency is much larger than the pulse spectral width, i.e. $\Delta \gg 1/\tau_r$. However, we can quantify adiabaticity by expanding in orders of $1/(\Delta \cdot \tau_r )$, which appears naturally in terms of the time derivatives of the pulse as
\begin{align}
\begin{split}
\int^{t}dt' \Omega(t') e^{i\Delta(t-t')} &= \frac{\Omega(t)}{-i\Delta}-\frac{\dot{\Omega}(t)}{(-i\Delta)^2}+\cdots\\
& = \sum\limits_{n=0}^{\infty} \frac{(-1)^n}{(-i\Delta)^{n+1}}\frac{d^n \Omega(t)}{dt^n}\;.
\end{split}
\label{Eq:SWPT-AdiabExp}
\end{align}
The right hand side of Eq.~(\ref{Eq:SWPT-AdiabExp}) is computed via integration by parts. Keeping the first term in the adiabatic expansion~(\ref{Eq:SWPT-AdiabExp}) agrees with the time-independent perturbation theory, while higher order terms capture the transient effects during pulse ramps.

Lastly, we note that the idea behind time-dependent SWPT and Magnus are similar, where a perturbative expansion is made in terms of the \textit{generator} (logarithm) of unitary operators. The two methods can be related to one another as
\begin{align}
\hat{U}_{\text{I}}(t,t_0)\equiv e^{-i \hat{K}(t,t_0)} = e^{-i\hat{G}(t)}\hat{U}_{\text{I,eff}}(t,t_0) e^{i\hat{G}(t_0)}
\label{Eq:SWPT-MagnusVsSWPT}
\end{align}
where $\hat{K}(t,t_0)$ is the Magnus generator. The main distinction is that Magnus solves for the overall time evolution operator \textit{directly}, without partitioning into resonant and off-resonant sectors.

In summary, we have demonstrated the application of time-dependent SWPT for computing an \textit{effective} Hamiltonian for a driven quantum system. This method is capable of accounting for the renormalization of effective Hamiltonian rates due to control pulse shapes, and also the corresponding off-resonant error.


\section{Effective CR Hamiltonian}
\label{App:EffCRHam}

Here, we apply the time-dependent SWPT of Appendix~\ref{App:SWPT} to the CR model in Eqs.~(\ref{eqn:DirCNOT-Def of H0})--(\ref{eqn:DirCNOT-Def of Hd(t)}) and provide expressions for the effective gate parameters. Our time-dependent results agree with and provide a natural extension of the time-independent CR rates given in Refs.~\cite{Tripathi_Operation_2019, Magesan_Effective_2020, Malekakhlagh_First-Principles_2020}.  

For CR gate, the drive frequency is resonant with the target qubit leading to Rabi oscillations around the X or Y axis of the target. Therefore, the frame in which the effective rates are resonant is BD with respect to the control qubit (see Fig.~\ref{fig:OffRes-BDFrame}). Once the effective Hamiltonian is obtained in the extended Hilbert space, we read off the CR gate parameters as
\begin{subequations}
\begin{align}
&\HO_{\text{CR,eff}}(t) \equiv \sum\limits_{m,n=i,x,y,z}\frac{1}{2}\omega_{\sigma_m \sigma_n }(t) \hat{\sigma}_m \otimes \hat{\sigma}_n \;,
\label{Eq:SWPT-Def of H_cR,eff}\\
&\omega_{\sigma_m \sigma_n} (t)\equiv \frac{1}{2}\Tr \left(\left(\hat{\sigma}_m \otimes \hat{\sigma}_n \right)\HO_{\text{I,eff}}(t)\right)  \;,
\label{Eq:SWPT-Def of w_(sigma)}
\end{align}
\end{subequations}
where the order is control $\otimes$ target. In the BD frame, the effective Hamiltonian consists of $IX$, $IY$, $ZX$, $ZY$, $IZ$, $ZI$ and $ZZ$ rates. In the following, we provide expressions in powers of drive amplitudes $\Omega_c(t)$ and $\Omega_t(t)$. 

\subsection{Zeroth order}
Up to the zeroth order, we find an effective static $ZZ$ rate as
\begin{align}
\omega_{zz}^{(0)}(t)= \left(\frac{1}{\Delta_{ct}-\alpha_t}-\frac{1}{\Delta_{ct}+\alpha_c}\right)J^2 +O\left(J^4\right)\;,
\label{Eq:EffCRHam-0thZZ}
\end{align} 
as a result of level repulsion between states $\ket{11}\leftrightarrow \ket{02}$ and $\ket{11}\leftrightarrow \ket{20}$.

\subsection{First order}
Up to the first order, we find corrections to the $IX$, $IY$, $ZX$ and $ZY$ rates as
\begin{subequations}
\begin{align}
&\omega_{ix}^{(1)}(t)=\Omega_{tx}(t)-\frac{J}{\Delta_{ct}+\alpha_c}\Omega_{cx}(t) \;,
\label{Eq:EffCRHam-1stIX}\\
&\omega_{iy}^{(1)}(t)=\Omega_{ty}(t)-\frac{J}{\Delta_{ct}+\alpha_c}\Omega_{cy}(t) \;,
\label{Eq:EffCRHam-1stIY}\\
&\omega_{zx}^{(1)}(t)=\left(\frac{J}{\Delta_{ct}+\alpha_c}-\frac{J}{\Delta_{ct}}\right)\Omega_{cx}(t) \;,
\label{Eq:EffCRHam-1stZX}\\
&\omega_{zy}^{(1)}(t)=\left(\frac{J}{\Delta_{ct}+\alpha_c}-\frac{J}{\Delta_{ct}}\right)\Omega_{cy}(t) \;.
\label{Eq:EffCRHam-1stZX}
\end{align}
\end{subequations}
In particular, $IX$ and $IY$ depend \textit{directly} on the resonant target drive, while the dependence on the control drive is \textit{indirect} and mediated through states $\ket{10}$ and $\ket{20}$, resulting in energy denominators $\Delta_{ct}$ and $\Delta_{ct}+\alpha_c$ (see Fig.~\ref{fig:CRSchematicPlusEnergyDiagram}).  


\subsection{Second order}
Up to the second order in drive amplitudes, there are corrections to the diagonal components $ZI$, $IZ$ and $ZZ$, proportional to $\Omega_{cx}^2+\Omega_{cy}^2$, $J^2(\Omega_{cx}^2+\Omega_{cy}^2)$, $J(\Omega_{cx}\Omega_{tx}+\Omega_{cy}\Omega_{ty})$, $\Omega_{cx}\dot{\Omega}_{cy}-\Omega_{cy}\dot{\Omega}_{cx}$, $J^2(\Omega_{cx}\dot{\Omega}_{cy}-\Omega_{cy}\dot{\Omega}_{cx})$ and $J(\Omega_{tx}\dot{\Omega}_{cy}-\Omega_{ty}\dot{\Omega}_{cx})$. For simplicity, we have performed the adiabtic expansion~(\ref{Eq:SWPT-AdiabExp}) up to the leading order in the pulse derivative. For instance, the expression for $\omega_{zi}^{(2)}(t)$ reads
\begin{subequations}
\begin{align}
\begin{split}
\omega_{zi}^{(2)}(t)&=C_{zi,1}^{(2)}\left[\Omega_{cx}^2(t)+\Omega_{cy}^2(t)\right]\\
&+C_{zi,2}^{(2)}J^2\left[\Omega_{cx}^2(t)+\Omega_{cy}^2(t)\right]\\
&+C_{zi,3}^{(2)}J\left[\Omega_{tx}(t)\Omega_{cx}(t)+\Omega_{ty}(t)\Omega_{cy}(t)\right]\\
&+C_{zi,4}^{(2)}\left[\Omega_{cx}(t)\dot{\Omega}_{cy}(t)-\Omega_{cy}(t)\dot{\Omega}_{cx}(t)\right]\\
&+C_{zi,5}^{(2)}J^2\left[\Omega_{cx}(t)\dot{\Omega}_{cy}(t)-\Omega_{cy}(t)\dot{\Omega}_{cx}(t)\right]\\
&+C_{zi,6}^{(2)}J\left[\Omega_{tx}(t)\dot{\Omega}_{cy}(t)-\Omega_{ty}(t)\dot{\Omega}_{cx}(t)\right] \;,
\end{split}
\label{Eq:EffCRHam-2ndZI}
\end{align}
\end{subequations}
where $C_{zi,1}^{(2)}$ to $C_{zi,6}^{(2)}$ are the corresponding energy denominators given in the following. Similar expressions for $\omega_{iz}^{(2)}(t)$ and $\omega_{zz}^{(2)}(t)$ follows.

For Stark shift on the control qubit, the energy denominators are found as
\begin{subequations}
\begin{align}
C_{zi,1}^{(2)}&=\frac{1}{2(\Delta_{ct}+\alpha_c)}-\frac{1}{2\Delta_{ct}}\;,
\label{Eq:EffCRHam-C_zi,1^(2)}
\end{align}
\begin{align}
\begin{split}
C_{zi,2}^{(2)}&=-\frac{1}{4 \alpha_c \Delta _{ct}^2}+\frac{1}{\alpha_c \left(\alpha_c+\Delta_{ct}\right){}^2}-\frac{3}{4 \alpha_c \left(2\alpha_c+\Delta_{ct}\right){}^2}\\
&+\frac{1}{\alpha_c^2 \Delta_{ct}}-\frac{2}{\alpha_c^2\left(\alpha_c+\Delta_{ct}\right)}-\frac{3}{\alpha_c^2 \left(2 \alpha_c+\Delta_{ct}\right)}\\
&-\frac{4}{\alpha_c^2\left(\alpha _c+2 \Delta_{ct}\right)}+\frac{12}{\alpha_c^2 \left(3 \alpha_c+2 \Delta_{ct}\right)}\\
&+\frac{1}{\alpha_t\left(\alpha _c+\Delta _{ct}-\alpha_t\right){}^2}-\frac{1}{\alpha_c \alpha_t \left(\Delta_{ct}-\alpha _t\right)}\\
&+\frac{1}{\alpha _c \alpha_t \left(\alpha_c+\Delta_{ct}-\alpha_t\right)} \;,
\end{split}
\label{Eq:EffCRHam-C_zi,2^(2)}
\end{align}
\begin{align}
C_{zi,3}^{(2)}&=\frac{\alpha_c}{2 \Delta_{ct} \alpha_t \left(\alpha_c+\Delta _{ct}\right)} \;,
\label{Eq:EffCRHam-C_zi,3^(2)}
\end{align}
\begin{align}
C_{zi,4}^{(2)}&=\frac{1}{2\Delta_{ct}^2}-\frac{1}{2(\Delta_{ct}+\alpha_c)^2	}\;,
\label{Eq:EffCRHam-C_zi,4^(2)}
\end{align}
\begin{align}
\begin{split}
C_{zi,5}^{(2)}&=\frac{6}{\alpha_c^3 \left(\alpha_c+\Delta_{ct}\right)}-\frac{9}{2 \alpha_c^3 \left(2 \alpha_c+\Delta_{ct}\right)}\\
&-\frac{1}{2 \alpha_c^2 \left(\alpha_c+\Delta_{ct}\right){}^2}-\frac{3}{4 \alpha_c^2 \left(2 \alpha_c+\Delta_{ct}\right){}^2}\\
&+\frac{4}{\alpha_c^2 \left(\alpha_c+2 \Delta_{ct}\right){}^2}-\frac{12}{\alpha_c^2\left(3 \alpha_c+2 \Delta_{ct}\right){}^2}\\
&-\frac{3}{2 \alpha_c^3 \Delta_{ct}}+\frac{1}{4 \alpha _c^2 \Delta_{ct}^2}+\frac{1}{\alpha_c \alpha_t^2\left(\Delta _{ct}-\alpha_t\right)}\\
&-\frac{1}{\alpha_c \alpha_t^2 \left(\alpha_c+\Delta_{ct}-\alpha_t\right)}-\frac{1}{\alpha_t^2 \left(\alpha_c+\Delta _{ct}-\alpha_t\right){}^2} \;,
\end{split}
\label{Eq:EffCRHam-C_zi,5^(2)}
\end{align}
\begin{align}
C_{zi,6}^{(2)}&=-\frac{\alpha_c^2 \Delta_{ct}+\alpha_c \Delta_{ct}^2+2 \alpha_c \Delta_{ct} \alpha_t+\alpha_c^2 \alpha_t}{2 \Delta _{ct}^2 \alpha_t^2 \left(\alpha_c+\Delta_{ct}\right){}^2} \;.
\label{Eq:EffCRHam-C_zi,6^(2)}
\end{align}
\end{subequations}

For Stark shift on the target qubit, i.e. $IZ$ rate, the energy denominators read
\begin{subequations}
\begin{align}
C_{iz,1}^{(2)}&=0 \;,
\label{Eq:EffCRHam-C_iz,1^(2)}
\end{align}
\begin{align}
\begin{split}
C_{iz,2}^{(2)}&=\frac{1}{4 \alpha_c \Delta_{ct}^2}+\frac{1}{2\alpha_c \left(\alpha_c+\Delta_{ct}\right){}^2}-\frac{3}{4 \alpha_c \left(2\alpha_c+\Delta_{ct}\right){}^2}\\
&-\frac{1}{\alpha_c^2 \Delta_{ct}}-\frac{4}{\alpha_c^2\left(\alpha_c+\Delta_{ct}\right)}-\frac{3}{\alpha_c^2 \left(2 \alpha_c+\Delta_{ct}\right)}\\
&+\frac{4}{\alpha_c^2\left(\alpha_c+2 \Delta_{ct}\right)}+\frac{12}{\alpha_c^2 \left(3\alpha_c+2 \Delta _{ct}\right)}\\
&+\frac{1}{\alpha_t \left(\alpha_c+\Delta_{ct}-\alpha_t\right){}^2}-\frac{1}{\alpha_c \alpha_t \left(\Delta_{ct}-\alpha_t\right)}\\
&+\frac{1}{\alpha_c \alpha_t \left(\alpha_c+\Delta_{ct}-\alpha_t\right)}+\frac{1}{2 \alpha_t \left(\Delta_{ct}-\alpha_t\right){}^2}\;,
\end{split}
\label{Eq:EffCRHam-C_iz,2^(2)}
\end{align}
\begin{align}
C_{iz,3}^{(2)}&= -\frac{\Delta_{ct}+\alpha_c+\alpha_t}{2 \alpha_t\left(\alpha_c+\Delta_{ct}\right){}^2} \;,
\label{Eq:EffCRHam-C_iz,3^(2)}
\end{align}
\begin{align}
C_{iz,4}^{(2)}&= 0\;,
\label{Eq:EffCRHam-C_iz,4^(2)}
\end{align}
\begin{align}
\begin{split}
C_{iz,5}^{(2)}&= \frac{3}{\alpha_c^3 \left(\alpha_c+\Delta_{ct}\right)}-\frac{9}{2 \alpha _c^3 \left(2 \alpha_c+\Delta_{ct}\right)}\\
&-\frac{1}{\alpha_c^2\left(\alpha_c+\Delta_{ct}\right){}^2}-\frac{3}{4 \alpha_c^2 \left(2 \alpha_c+\Delta_{ct}\right){}^2}\\
&-\frac{4}{\alpha_c^2 \left(\alpha_c+2 \Delta_{ct}\right){}^2}-\frac{12}{\alpha_c^2 \left(3 \alpha_c+2 \Delta_{ct}\right){}^2}\\
&+\frac{3}{2 \alpha_c^3 \Delta_{ct}}-\frac{1}{4 \alpha_c^2 \Delta_{ct}^2}+\frac{1}{\alpha_c \alpha_t^2 \left(\Delta_{ct}-\alpha_t\right)}\\
&-\frac{1}{\alpha_c \alpha_t^2 \left(\alpha_c+\Delta_{ct}-\alpha_t\right)}-\frac{1}{2 \alpha_t^2 \left(\Delta_{ct}-\alpha_t\right){}^2}\\
&-\frac{1}{\alpha_t^2 \left(\alpha_c+\Delta_{ct}-\alpha_t\right){}^2}\;,
\end{split}
\label{Eq:EffCRHam-C_iz,5^(2)}
\end{align}
\begin{align}
C_{iz,6}^{(2)}&=\frac{\alpha_c^2+2 \alpha_c \Delta_{ct}+\alpha_c\alpha_t+\Delta_{ct}^2+\Delta_{ct} \alpha_t+\alpha_t^2}{2 \alpha_t^2 \left(\alpha_c+\Delta_{ct}\right){}^3}\;.
\label{Eq:EffCRHam-C_iz,6^(2)}
\end{align}
\end{subequations}

Similarly, for the dynamic $ZZ$ rate, the energy denominators read
\begin{subequations}
\begin{align}
C_{zz,1}^{(2)}&= 0\;,
\label{Eq:EffCRHam-C_zz,1^(2)}\\
\begin{split}
C_{zz,2}^{(2)}&= \frac{1}{4 \alpha_c \Delta_{\text{ct}}^2}-\frac{1}{\alpha_c \left(\alpha_c+\Delta_{\text{ct}}\right){}^2}+\frac{3}{4 \alpha_c \left(2\alpha_c+\Delta_{\text{ct}}\right){}^2}\\
&-\frac{1}{\alpha_c^2 \Delta_{\text{ct}}}+\frac{2}{\alpha_c^2 \left(\alpha_c+\Delta_{\text{ct}}\right)}+\frac{3}{\alpha_c^2 \left(2 \alpha_c+\Delta_{\text{ct}}\right)}\\
&+\frac{4}{\alpha_c^2 \left(\alpha _c+2 \Delta_{\text{ct}}\right)}-\frac{12}{\alpha_c^2 \left(3 \alpha_c+2 \Delta_{\text{ct}}\right)}\\
&-\frac{1}{\alpha_t \left(\alpha_c+\Delta_{\text{ct}}-\alpha_t\right){}^2}+\frac{1}{\alpha_c \alpha_t \left(\Delta_{\text{ct}}-\alpha_t\right)}\\
&-\frac{1}{\alpha_c \alpha_t \left(\alpha_c+\Delta_{\text{ct}}-\alpha_t\right)}\;,
\end{split}
\label{Eq:EffCRHam-C_zz,2^(2)}\\
C_{zz,3}^{(2)}&=-\frac{\alpha_c^2 \Delta_{\text{ct}}+\alpha_c \Delta_{\text{ct}}^2+2 \alpha_c \Delta_{\text{ct}} \alpha_t+\alpha_c^2 \alpha_t}{2 \Delta_{\text{ct}}^2 \alpha_t \left(\alpha_c+\Delta_{\text{ct}}\right){}^2}\;,
\label{Eq:EffCRHam-C_zz,3^(2)}\\
C_{zz,4}^{(2)}&= 0\;,
\label{Eq:EffCRHam-C_zz,4^(2)}\\
\begin{split}
C_{zz,5}^{(2)}&= -\frac{6}{\alpha_c^3 \left(\alpha_c+\Delta_{\text{ct}}\right)}+\frac{9}{2 \alpha_c^3 \left(2 \alpha_c+\Delta_{\text{ct}}\right)}\\
&+\frac{1}{2\alpha_c^2\left(\alpha_c+\Delta_{\text{ct}}\right){}^2}+\frac{3}{4\alpha_c^2 \left(2 \alpha_c+\Delta_{\text{ct}}\right){}^2}\\
&-\frac{4}{\alpha_c^2 \left(\alpha_c+2 \Delta_{\text{ct}}\right){}^2}+\frac{12}{\alpha_c^2\left(3 \alpha_c+2 \Delta_{\text{ct}}\right){}^2}\\
&+\frac{3}{2 \alpha_c^3 \Delta_{\text{ct}}}-\frac{1}{4 \alpha_c^2 \Delta_{\text{ct}}^2}-\frac{1}{\alpha_c \alpha_t^2 \left(\Delta_{\text{ct}}-\alpha_t\right)}\\
&+\frac{1}{\alpha_c \alpha_t^2 \left(\alpha_c+\Delta_{\text{ct}}-\alpha_t\right)}+\frac{1}{\alpha_t^2 \left(\alpha_c+\Delta_{\text{ct}}-\alpha_t\right){}^2}\;,
\end{split}
\label{Eq:EffCRHam-C_zz,5^(2)}
\end{align}
\begin{align}
\begin{split}
C_{zz,6}^{(2)}&= -\frac{1}{2 \left(\alpha_c+\Delta_{\text{ct}}\right){}^3}-\frac{1}{2 \alpha_t \left(\alpha_c+\Delta_{\text{ct}}\right){}^2}\\
&-\frac{1}{2 \alpha_t^2 \left(\alpha_c+\Delta_{\text{ct}}\right)}+\frac{1}{2\Delta_{\text{ct}}^3}+\frac{1}{2 \Delta_{\text{ct}}^2\alpha_t}+\frac{1}{2 \Delta_{\text{ct}} \alpha_t^2} \;.
\end{split}
\label{Eq:EffCRHam-C_zz,6^(2)}
\end{align}
\end{subequations}

\subsection{Third order}
Up to the third order in drive amplitudes, we find corrections to the $IX$, $IY$, $ZX$ and $ZY$ rates. Here, there are numerous contributions and, for brevity, we quote certain dominant corrections to e.g. the $IX$ as  
\begin{align}
\begin{split}
\omega_{ix}^{(3)}(t)&= C_{ix,1}^{(3)} J\left[\Omega_{cx}^2(t)+\Omega_{cy}^2(t)\right]\Omega_{cx}(t)\\
&+C_{ix,2}^{(3)} J \dot{\Omega}_{cx}^2(t) \Omega_{cx}(t)\\
&+C_{ix,3}^{(3)} J\dot{\Omega}_{cy}^2(t) \Omega_{cx}(t) \;.
\end{split}
\label{Eq:EffCRHam-3rdIX}
\end{align}
Expressions for the $ZX$, $IY$, and $ZY$ rates have a similar form.

The energy denominators for the $IX$ rate read	
\begin{subequations}
\begin{align}
C_{ix,1}^{(3)} & = \frac{\alpha_c \Delta_{\text{ct}}}{\left(\alpha_c+\Delta_{\text{ct}}\right){}^3 \left(\alpha_c+2 \Delta_{\text{ct}}\right) \left(3 \alpha_c+2 \Delta_{\text{ct}}\right)} \;, 
\label{Eq:EffCRHam-C_ix,1^(3)}
\end{align}
\begin{align}
\begin{split}
C_{ix,2}^{(3)} & = -\frac{1}{\left(\alpha _c+\Delta_{\text{ct}}\right){}^5}+\frac{1}{2 \alpha_c \left(\alpha_c+\Delta_{\text{ct}}\right){}^4}\\
&+\frac{1}{12 \alpha _c^2 \Delta_{\text{ct}}^3}-\frac{5}{24 \alpha_c^2 \left(\alpha_c+\Delta_{\text{ct}}\right){}^3}-\frac{1}{3 \alpha_c^3 \Delta_{\text{ct}}^2}\\
&-\frac{71}{24 \alpha_c^3 \left(\alpha_c+\Delta_{\text{ct}}\right){}^2}+\frac{2}{3 \alpha_c^3 \left(\alpha_c+2 \Delta_{\text{ct}}\right){}^2}\\
&-\frac{12}{\alpha_c^3 \left(3 \alpha_c+2 \Delta_{\text{ct}}\right){}^2}+\frac{103}{8 \alpha _c^4 \left(\alpha_c+\Delta_{\text{ct}}\right)}\\
&+\frac{9}{8 \alpha_c^4 \Delta _{\text{ct}}}-\frac{4}{\alpha_c^4 \left(\alpha_c+2 \Delta_{\text{ct}}\right)}-\frac{24}{\alpha_c^4\left(3 \alpha _c+2 \Delta _{\text{ct}}\right)} 
\end{split}
\label{Eq:EffCRHam-C_ix,2^(3)}
\end{align}
\begin{align}
C_{ix,3}^{(3)} & = \frac{\alpha_c \left(4 \alpha_c+7 \Delta_{\text{ct}}\right)}{24 \Delta _{\text{ct}}^2 \left(\alpha_c+\Delta_{\text{ct}}\right){}^3\left(\alpha_c+2 \Delta_{\text{ct}}\right){}^2}\;,
\label{Eq:EffCRHam-C_ix,3^(3)}
\end{align}
\end{subequations}
and for the $ZX$ rate are found as
\begin{subequations}
\begin{align}
C_{zx,1}^{(3)} & = \frac{\alpha_c^2 \left(3 \alpha_c^3+11 \alpha_c^2 \Delta_{\text{ct}}+15 \alpha_c \Delta_{\text{ct}}^2+9 \Delta_{\text{ct}}^3\right)}{2\Delta_{\text{ct}}^3 \left(\alpha_c+\Delta_{\text{ct}}\right){}^3 \left(\alpha_c+2 \Delta_{\text{ct}}\right) \left(3\alpha_c+2 \Delta_{\text{ct}}\right)} ,
\label{Eq:EffCRHam-C_zx,1^(3)}
\end{align}
\begin{align}
\begin{split}
C_{zx,2}^{(3)} & = \frac{1}{\left(\alpha_c+\Delta_{\text{ct}}\right){}^5}+\frac{1}{2 \alpha_c \Delta _{\text{ct}}^4}-\frac{1}{2 \alpha_c \left(\alpha_c+\Delta_{\text{ct}}\right){}^4}\\
&-\frac{1}{2 \alpha_c^2\Delta _{\text{ct}}^3}+\frac{3}{8 \alpha_c^2 \left(\alpha_c+\Delta_{\text{ct}}\right){}^3}+\frac{1}{4 \alpha_c^3 \Delta_{\text{ct}}^2}\\
&+\frac{29}{8 \alpha_c^3\left(\alpha_c+\Delta_{\text{ct}}\right){}^2}-\frac{2}{\alpha_c^3 \left(\alpha_c+2 \Delta_{\text{ct}}\right){}^2}\\
&+\frac{12}{\alpha_c^3 \left(3 \alpha_c+2 \Delta_{\text{ct}}\right){}^2}+\frac{5}{8 \alpha_c^4 \Delta_{\text{ct}}}-\frac{85}{8 \alpha_c^4 \left(\alpha_c+\Delta_{\text{ct}}\right)}\\
&-\frac{4}{\alpha _c^4 \left(\alpha_c+2 \Delta_{\text{ct}}\right)}+\frac{24}{\alpha_c^4 \left(3 \alpha_c+2 \Delta_{\text{ct}}\right)}-\frac{1}{2 \Delta_{\text{ct}}^5}\;,
\end{split}
\label{Eq:EffCRHam-C_zx,2^(3)}
\end{align}
\begin{align}
C_{zx,3}^{(3)} & = -\frac{\alpha_c \left(2 \alpha_c^2+8 \alpha_c \Delta_{\text{ct}}+7 \Delta_{\text{ct}}^2\right)}{8 \Delta_{\text{ct}}^3 \left(\alpha_c+\Delta_{\text{ct}}\right){}^3 \left(\alpha_c+2 \Delta_{\text{ct}}\right){}^2}\;.
\label{Eq:EffCRHam-C_zx,3^(3)}
\end{align}
\end{subequations}

\subsection{fourth order}
There are multitude of corrections up to the fourth order. Among those, here, we quote the most dominant contribution to the control qubit Stark shift as 
\begin{align}
\begin{split}
\omega_{zi}^{(4)}(t) & \equiv \frac{\left(3 \alpha_c^5+ 11\alpha_c^4\Delta_{ct}+15 \alpha_c^3 \Delta_{ct}^2+ 9\alpha_c^2 \Delta_{ct}^3\right)}{8 \Delta _{ct}^3 \left(\alpha_c+ \Delta_{ct}\right)^3 \left(\alpha_c+ 2 \Delta_{ct}\right)\left(3 \alpha_c+2 \Delta_{ct}\right)} \\
& \times \left[\Omega_{cx}^2(t)+\Omega_{cy}^2(t)\right]^2 \;.
\end{split}
\label{Eq:EffCRHam-C_zx,3^(3)}   
\end{align}
Accounting for Eq.~(\ref{Eq:EffCRHam-C_zx,3^(3)}), on top of the 1st line of Eq.~(\ref{Eq:EffCRHam-2ndZI}), becomes important at stronger CR drive which suppresses the Stark shift in magnitude [see also Fig.~(3c) of Ref.~\cite{Malekakhlagh_First-Principles_2020}].   

In summary, our time-dependent formalism captures the previously known expressions for the gate parameters, e.g. Eq.~(\ref{Eq:EffCRHam-C_zi,1^(2)}) for $ZI$, Eq.~(\ref{Eq:EffCRHam-C_ix,1^(3)}) for $IX$ and Eq.~(\ref{Eq:EffCRHam-C_zx,1^(3)}) for $ZX$, while also quantifies non-adiabatic response in terms of the derivative of underlying pulse shapes.

\section{Effective CR time evolution operator}
\label{App:EffU}
Here, we calculate the \textit{effective} time evolution operator for the CR gate as
\begin{align}
\hat{U}_{\text{CR,eff}}(t,0) = \mathbb{T} \exp \left[-i \int_{0} ^{t} dt' \HO_{\text{CR,eff}}(t')\right] \;,
\label{Eq:EffU-Def of U_I,eff(t,t0)}
\end{align}
where $\HO_{\text{CR,eff}}(t)$ is the effective BD Hamiltonian in Appendix~\ref{App:EffCRHam}. In general, time-dependent corrections in the effective rates due to pulse ramps do not commute, hence explicit time-ordering is needed. For simplicity, however, we keep the leading order in the adiabatic expansion which corresponds to the constant mid-part of the pulses. The resulting approximate expressions are helpful for designing specific gate calibrations and reverse engineering the required drive scheme \cite{Malekakhlagh_First-Principles_2020, Sundaresan_Reducing_2020}. 

In the control=$\ket{0}$ subspace, the $\ket{00}\bra{00}$, $\ket{00}\bra{01}$, $\ket{01}\bra{00}$ and $\ket{01}\bra{01}$ components of $\hat{U}_{\text{CR,eff}}(t,0)$ are found respectively as
\begin{subequations}
\begin{align}
&\frac{e^{-\frac{1}{2} i \omega _{zi} t}\left[\omega _+ \cos \left(\frac{\omega_{+} t}{2}\right)-i \left(\omega _{iz}+\omega_{zz}\right) \sin \left(\frac{\omega_{+} t}{2}\right)\right]}{\omega _+}\;,
\label{Eq:EffU-u00,00}\\
&-\frac{e^{-\frac{1}{2}i \omega_{zi} t}\left[i(\omega_{ix}+\omega_{zx})+(\omega_{iy}+\omega_{zy})\right]\sin\left(\frac{\omega_{+} t}{2}\right)}{\omega_{+}}\;, 
\label{Eq:EffU-u00,01}\\
&-\frac{e^{-\frac{1}{2}i \omega_{zi} t}\left[i(\omega_{ix}+\omega_{zx})-(\omega_{iy}+\omega_{zy})\right]\sin\left(\frac{\omega_{+} t}{2}\right)}{\omega_{+}} \;,
\label{Eq:EffU-u01,00}\\
&\frac{e^{-\frac{1}{2} i \omega _{zi} t}\left[\omega _+ \cos \left(\frac{\omega_{+} t}{2}\right)+i \left(\omega _{iz}+\omega_{zz}\right) \sin \left(\frac{\omega_{+} t}{2}\right)\right]}{\omega _+} \;,
\label{Eq:EffU-u01,01}
\end{align}
\end{subequations}
where $\omega_{\pm}$ are collective CR frequencies \cite{Malekakhlagh_First-Principles_2020, Sundaresan_Reducing_2020} defined as
\begin{align}
\omega_{\pm}\equiv \left[(\omega_{ix}\pm\omega_{zx})^2+(\omega_{iy}\pm\omega_{zy})^2+(\omega_{iz}\pm\omega_{zz})^2\right]^{1/2} \;.
\label{Eq:EffU-Def of w_pm}
\end{align}
Similarly, in the control=$\ket{1}$ subspace, the $\ket{10}\bra{10}$, $\ket{10}\bra{11}$, $\ket{11}\bra{10}$ and $\ket{11}\bra{11}$ components of $\hat{U}_{\text{CR,eff}}(t,0)$ read 
\begin{subequations}
\begin{align}
&\frac{e^{\frac{1}{2} i \omega _{zi} t}\left[\omega_{-} \cos \left(\frac{\omega_{-} t}{2}\right)-i \left(\omega _{iz}-\omega_{zz}\right) \sin \left(\frac{\omega_{-} t}{2}\right)\right]}{\omega _-} \;,
\label{Eq:EffU-u10,10}\\
&-\frac{e^{\frac{1}{2}i \omega_{zi} t}\left[i(\omega_{ix}-\omega_{zx})+(\omega_{iy}-\omega_{zy})\right]\sin\left(\frac{\omega_{-} t}{2}\right)}{\omega_{-}} \;,
\label{Eq:EffU-u10,11}\\
&-\frac{e^{\frac{1}{2}i \omega_{zi} t}\left[i(\omega_{ix}-\omega_{zx})-(\omega_{iy}-\omega_{zy})\right]\sin\left(\frac{\omega_{-} t}{2}\right)}{\omega_{-}} \;, 
\label{Eq:EffU-u11,10}\\
&\frac{e^{\frac{1}{2} i \omega _{zi} t}\left[\omega_{-} \cos \left(\frac{\omega_{-} t}{2}\right)+i \left(\omega _{iz}-\omega_{zz}\right) \sin \left(\frac{\omega_{-} t}{2}\right)\right]}{\omega_{-}} \;.
\label{Eq:EffU-u11,11}
\end{align}
\end{subequations}
According to Eq.~(\ref{Eq:EffU-u00,00})--(\ref{Eq:EffU-u11,11}), to calibrate a direct CNOT gate based on cross-resonance, it is needed to set $\omega_{ix}(t)+\omega_{zx}(t)=0$ and tune $\int_{0}^{\tau_p} dt'[\omega_{ix}(t')-\omega_{zx}(t')]=\pi$ as a $\pi$ pulse. These conditions along with perturbative estimates for gate parameters in Appendix~\ref{App:EffCRHam} lead to the approximate calibration conditions~(\ref{eqn:DirCNOT-PertCond IX+ZX=0}) and~(\ref{eqn:DirCNOT-PertCond IX-ZX=Pi}) of the main text.  

Alternatively, we can express the effective unitary in the two-qubit Pauli basis as $u_{\text{eff},\sigma_m\sigma_n}(t,0) \equiv (1/4)\Tr \left\{(\hat{\sigma}_m \otimes \hat{\sigma}_n) \hat{U}_{\text{CR,eff}}(t,0) \right\}$. The Pauli decomposition reads
\begin{subequations}
\begin{align}
\begin{split}
u_{\text{eff},ii}(t,0)&= \frac{1}{2}\left[e^{\frac{1}{2} i \omega_{zi} t} \cos\left(\frac{\omega_{-}t}{2}\right) \right. \\
& \left. +e^{-\frac{1}{2} i \omega_{zi} t} \cos\left(\frac{\omega_{+}t}{2}\right)\right]\;,
\end{split}
\label{Eq:EffU-uii}
\end{align}
\begin{align}
\begin{split}
u_{\text{eff},ix}(t,0)&=-i \left[\frac{\omega_{+}(\omega_{ix}-\omega_{zx})e^{\frac{1}{2}i\omega_{zi}t}\sin\left(\frac{\omega_{-}t}{2}\right)}{2\omega_{+}\omega_{-}}\right.\\
&+ \left. \frac{\omega_{-}(\omega_{ix}+\omega_{zx})e^{-\frac{1}{2}i\omega_{zi}t}\sin\left(\frac{\omega_{+}t}{2}\right)}{2\omega_{+}\omega_{-}} \right]\;,
\end{split}
\label{Eq:EffU-uix}
\end{align}
\begin{align}
\begin{split}
u_{\text{eff},iy}(t,0)&=-i \left[\frac{\omega_{+}(\omega_{iy}-\omega_{zy})e^{\frac{1}{2}i\omega_{zi}t}\sin\left(\frac{\omega_{-}t}{2}\right)}{2\omega_{+}\omega_{-}}\right.\\
&+ \left. \frac{\omega_{-}(\omega_{iy}+\omega_{zy})e^{-\frac{1}{2}i\omega_{zi}t}\sin\left(\frac{\omega_{+}t}{2}\right)}{2\omega_{+}\omega_{-}} \right]\;,
\end{split}
\label{Eq:EffU-uiy}
\end{align}
\begin{align}
\begin{split}
u_{\text{eff},iz}(t,0)&=-i \left[\frac{\omega_{+}(\omega_{iz}-\omega_{zz})e^{\frac{1}{2}i\omega_{zi}t}\sin\left(\frac{\omega_{-}t}{2}\right)}{2\omega_{+}\omega_{-}}\right.\\
&+ \left. \frac{\omega_{-}(\omega_{iz}+\omega_{zz})e^{-\frac{1}{2}i\omega_{zi}t}\sin\left(\frac{\omega_{+}t}{2}\right)}{2\omega_{+}\omega_{-}} \right]\;,
\end{split}
\label{Eq:EffU-uix}
\end{align}
\begin{align}
\begin{split}
u_{\text{eff},zi}(t,0)&=\frac{1}{2}\left[e^{-\frac{1}{2} i \omega_{zi} t} \cos\left(\frac{\omega_{+}t}{2}\right)\right.\\
&\left. -e^{\frac{1}{2} i \omega_{zi} t} \cos\left(\frac{\omega_{-}t}{2}\right)\right]\;,
\end{split}
\label{Eq:EffU-uzi}
\end{align}
\begin{align}
\begin{split}
u_{\text{eff},zx}(t,0)&= i \left[\frac{\omega_{+}(\omega_{ix}-\omega_{zx})e^{\frac{1}{2}i\omega_{zi}t}\sin\left(\frac{\omega_{-}t}{2}\right)}{2\omega_{+}\omega_{-}}\right.\\
&-\left. \frac{\omega_{-}(\omega_{ix}+\omega_{zx})e^{-\frac{1}{2}i\omega_{zi}t}\sin\left(\frac{\omega_{+}t}{2}\right)}{2\omega_{+}\omega_{-}} \right]\;,
\end{split}
\label{Eq:EffU-uzx}
\end{align}
\begin{align}
\begin{split}
u_{\text{eff},zy}(t,0) &= i \left[\frac{\omega_{+}(\omega_{iy}-\omega_{zy})e^{\frac{1}{2}i\omega_{zi}t}\sin\left(\frac{\omega_{-}t}{2}\right)}{2\omega_{+}\omega_{-}}\right.\\
&-\left. \frac{\omega_{-}(\omega_{iy}+\omega_{zy})e^{-\frac{1}{2}i\omega_{zi}t}\sin\left(\frac{\omega_{+}t}{2}\right)}{2\omega_{+}\omega_{-}} \right]\;,
\end{split}
\label{Eq:EffU-uzy}
\end{align}
\begin{align}
\begin{split}
u_{\text{eff},zz}(t,0)&= i \left[\frac{\omega_{+}(\omega_{iz}-\omega_{zz})e^{\frac{1}{2}i\omega_{zi}t}\sin\left(\frac{\omega_{-}t}{2}\right)}{2\omega_{+}\omega_{-}}\right.\\
&-\left. \frac{\omega_{-}(\omega_{iz}+\omega_{zz})e^{-\frac{1}{2}i\omega_{zi}t}\sin\left(\frac{\omega_{+}t}{2}\right)}{2\omega_{+}\omega_{-}} \right]\;.
\end{split}
\label{Eq:EffU-uzz}
\end{align}
\end{subequations}

In summary, based on Eqs.~(\ref{Eq:EffU-u00,00})--(\ref{Eq:EffU-uzz}), the time evolution operator in the BD frame consists of beatings between three CR frequencies: $\omega_{+}$, $\omega_{-}$ and $\omega_{zi}$. 

\section{Non-BD terms in overall time evolution operator}
\label{App:NonBDU}

The BD decomposition of the effective time evolution operator, found in Appendix~\ref{App:EffU}, contains only the effective interactions. As discussed in Appendix~\ref{App:SWPT}, we can also quantify the off-resonant contributions by mapping the effective unitary operator back to the initial interaction frame as
\begin{align}
\hat{U}_{I}(t,0)=\hat{U}_{\text{SW}}(t)\hat{U}_{I,\text{eff}}(t,0)\hat{U}_{\text{SW}}^{\dag}(0) \;.
\label{Eq:NonBD-U_I(t,t0) ITO U_I,eff(t,t0)}
\end{align}
We then read off $\hat{U}_{\text{CR}}(t,0)$ as the projection of $\hat{U}_{I}(t,0)$ onto the computational subspace. Upon the frame transformation~(\ref{Eq:NonBD-U_I(t,t0) ITO U_I,eff(t,t0)}), both the effective BD and non-BD subspaces of $\hat{U}_{I}(t,0)$ are renormalized, while the corrections to the BD subspace is higher order. Here, we provide the lowest order adiabatic expressions for the non-BD elements of $\hat{U}_{\text{CR}}(t,0)$.

We begin by the lowest order expression for the $XI$ element of $\hat{U}_{\text{CR}}(t,0)$ as
\begin{align}
\begin{split}
u_{xi}(t,0) &= \frac{\Omega_c^*(0)+e^{-i\Delta_{ct} t}\Omega_c^*(t)}{4\Delta_{ct}} e^{\frac{1}{2}i\omega_{zi}t}\cos \left(\frac{\omega_{-}t}{2}\right) \\
&-\frac{\Omega_c(0)+e^{i\Delta_{ct} t}\Omega_c(t)}{4\Delta_{ct}} e^{-\frac{1}{2}i\omega_{zi}t}\cos \left(\frac{\omega_{+}t}{2}\right) \\
&+i\frac{\dot{\Omega}_c^*(0)-e^{-i\Delta_{ct} t}\dot{\Omega}_c^*(t)}{4\Delta_{ct}^2} e^{\frac{1}{2}i\omega_{zi}t}\cos \left(\frac{\omega_{-}t}{2}\right) \\
&+i\frac{\dot{\Omega}_c(0)-e^{i\Delta_{ct} t}\dot{\Omega}_c(t)}{4\Delta_{ct}^2} e^{-\frac{1}{2}i\omega_{zi}t}\cos \left(\frac{\omega_{+}t}{2}\right) \;,
\end{split}
\label{Eq:NonBD-uxi(t)}
\end{align}
truncated up to the 1st order derivative $\dot{\Omega}_c(t)$. Compared to the BD part of the time evolution, there exist \textit{faster} oscillation in terms of qubit-qubit detuning $\Delta_{ct}$ on top of effective slower oscillations characterized by $\omega_{zi}$ and $\omega_{\pm}$. Given that the drive amplitude is set to zero at $t=0,\tau_p$, we find that the $XI$ error at $t=\tau_p$ is determined by the last two-terms as 
\begin{align}
\begin{split}
u_{xi}(\tau_p,0) &= i\frac{\dot{\Omega}_c^*(0)-e^{-i\Delta_{ct} \tau_p}\dot{\Omega}_c^*(\tau_p)}{4\Delta_{ct}^2} e^{\frac{1}{2}i\omega_{zi}\tau_p}\cos \left(\frac{\omega_{-}\tau_p}{2}\right) \\
&+i\frac{\dot{\Omega}_c(0)-e^{i\Delta_{ct} \tau_p}\dot{\Omega}_c(\tau_p)}{4\Delta_{ct}^2} e^{-\frac{1}{2}i\omega_{zi}\tau_p}\cos \left(\frac{\omega_{+}\tau_p}{2}\right) \;.
\end{split}
\label{Eq:NonBD-uxi(tau)}
\end{align}
Based on Eq.~(\ref{Eq:NonBD-uxi(tau)}), the lowest order $XI$ error is determined by $\dot{\Omega}_c(t)|_{t=0,\tau_p}/\Delta_{ct}^2$ and can be mitigated by (i) larger qubit-qubit detuning and (ii) smoother ramps.  

Similar expressions can be obtained for other non-BD components. For instance, the $XX$, $XY$, and $XZ$ components of $\hat{U}_{\text{CR}}(\tau_p,0)$ read
\begin{align}
\begin{split}
u_{xx}(\tau_p,0) &= \frac{(\omega_{ix}-\omega_{zx})\left[\dot{\Omega}_c^*(0)-e^{-i\Delta_{ct} \tau}\dot{\Omega}_c^*(\tau_p)\right]}{4\Delta_{ct}^2\omega_{-}}\\
&\times e^{\frac{1}{2}i\omega_{zi}\tau}\sin \left(\frac{\omega_{-}\tau_p}{2}\right) \\
&+\frac{(\omega_{ix}+\omega_{zx})\left[\dot{\Omega}_c(0)-e^{i\Delta_{ct} \tau_p}\dot{\Omega}_c(\tau_p)\right]}{4\Delta_{ct}^2\omega_{+}}\\
&\times e^{-\frac{1}{2}i\omega_{zi}\tau_p}\sin \left(\frac{\omega_{+}\tau_p}{2}\right)\;, 
\end{split}
\label{Eq:NonBD-uxx(tau)}
\end{align}
\begin{align}
\begin{split}
u_{xy}(\tau_p,0) &= \frac{(\omega_{iy}-\omega_{zy})\left[\dot{\Omega}_c^*(0)-e^{-i\Delta_{ct} \tau_p}\dot{\Omega}_c^*(\tau_p)\right]}{4\Delta_{ct}^2\omega_{-}}\\
&\times e^{\frac{1}{2}i\omega_{zi}\tau_p}\sin \left(\frac{\omega_{-}\tau_p}{2}\right) \\
&+\frac{(\omega_{iy}+\omega_{zy})\left[\dot{\Omega}_c(0)-e^{i\Delta_{ct} \tau_p}\dot{\Omega}_c(\tau_p)\right]}{4\Delta_{ct}^2\omega_{+}}\\
&\times e^{-\frac{1}{2}i\omega_{zi}\tau_p}\sin \left(\frac{\omega_{+}\tau_p}{2}\right) \;,
\end{split}
\label{Eq:NonBD-uxy(tau)}
\end{align}  
\begin{align}
\begin{split}
u_{xz}(\tau_p,0) &= \frac{(\omega_{iz}-\omega_{zz})\left[\dot{\Omega}_c^*(0)-e^{-i\Delta_{ct} \tau_p}\dot{\Omega}_c^*(\tau_p)\right]}{4\Delta_{ct}^2\omega_{-}}\\
&\times e^{\frac{1}{2}i\omega_{zi}\tau_p}\sin \left(\frac{\omega_{-}\tau_p}{2}\right) \\
&+\frac{(\omega_{iz}+\omega_{zz})\left[\dot{\Omega}_c(0)-e^{i\Delta_{ct} \tau_p}\dot{\Omega}_c(\tau_p)\right]}{4\Delta_{ct}^2\omega_{+}}\\
&\times e^{-\frac{1}{2}i\omega_{zi}\tau}\sin \left(\frac{\omega_{+}\tau_p}{2}\right) \;.
\end{split}
\label{Eq:NonBD-uxz(tau)}
\end{align} 
Furthermore, the $YI$, $YX$, $YY$ and $YZ$ components of $\hat{U}_{\text{CR}}(\tau_p,0)$ read
\begin{align}
\begin{split}
u_{yi}(\tau_p,0) &= \frac{\dot{\Omega}_c^*(0)+e^{-i\Delta_{ct} \tau_p}\dot{\Omega}_c^*(\tau_p)}{4\Delta_{ct}^2} e^{\frac{1}{2}i\omega_{zi}\tau_p}\cos \left(\frac{\omega_{-}\tau_p}{2}\right) \\
&-\frac{\dot{\Omega}_c(0)+e^{i\Delta_{ct} \tau_p}\dot{\Omega}_c(\tau_p)}{4\Delta_{ct}^2} e^{-\frac{1}{2}i\omega_{zi}\tau_p}\cos \left(\frac{\omega_{+}\tau_p}{2}\right)\;,
\end{split}
\label{Eq:NonBD-uyi(tau)}
\end{align}
\begin{align}
\begin{split}
u_{yx}(\tau_p,0) &= -i \frac{(\omega_{ix}-\omega_{zx})\left[\dot{\Omega}_c^*(0)+e^{-i\Delta_{ct} \tau_p}\dot{\Omega}_c^*(\tau_p)\right]}{4\Delta_{ct}^2\omega_{-}}\\
&\times e^{\frac{1}{2}i\omega_{zi}\tau_p}\sin \left(\frac{\omega_{-}\tau_p}{2}\right) \\
&+i \frac{(\omega_{ix}+\omega_{zx})\left[\dot{\Omega}_c(0)+e^{i\Delta_{ct} \tau_p}\dot{\Omega}_c(\tau_p)\right]}{4\Delta_{ct}^2\omega_{+}}\\
&\times e^{-\frac{1}{2}i\omega_{zi}\tau_p}\sin \left(\frac{\omega_{+}\tau_p}{2}\right) \;,
\end{split}
\label{Eq:NonBD-uyx(tau)}
\end{align}
\begin{align}
\begin{split}
u_{yy}(\tau_p,0) &= -i \frac{(\omega_{iy}-\omega_{zy})\left[\dot{\Omega}_c^*(0)+e^{-i\Delta_{ct} \tau}\dot{\Omega}_c^*(\tau_p)\right]}{4\Delta_{ct}^2\omega_{-}}\\
&\times e^{\frac{1}{2}i\omega_{zi}\tau}\sin \left(\frac{\omega_{-}\tau_p}{2}\right) \\
&+i \frac{(\omega_{iy}+\omega_{zy})\left[\dot{\Omega}_c(0)+e^{i\Delta_{ct} \tau}\dot{\Omega}_c(\tau_p)\right]}{4\Delta_{ct}^2\omega_{+}}\\
&\times e^{-\frac{1}{2}i\omega_{zi}\tau_p}\sin \left(\frac{\omega_{+}\tau_p}{2}\right)\;,
\end{split}
\label{Eq:NonBD-uyy(tau)}
\end{align}
\begin{align}
\begin{split}
u_{yz}(\tau_p,0) &= -i \frac{(\omega_{iz}-\omega_{zz})\left[\dot{\Omega}_c^*(0)+e^{-i\Delta_{ct} \tau_p}\dot{\Omega}_c^*(\tau_p)\right]}{4\Delta_{ct}^2\omega_{-}}\\
&\times e^{\frac{1}{2}i\omega_{zi}\tau_p}\sin \left(\frac{\omega_{-}\tau_p}{2}\right) \\
&+i \frac{(\omega_{iz}+\omega_{zz})\left[\dot{\Omega}_c(0)+e^{i\Delta_{ct} \tau}\dot{\Omega}_c(\tau_p)\right]}{4\Delta_{ct}^2\omega_{+}}\\
&\times e^{-\frac{1}{2}i\omega_{zi}\tau_p}\sin \left(\frac{\omega_{+}\tau_p}{2}\right)\;.
\end{split}
\label{Eq:NonBD-uyz(tau)}
\end{align}

Equations~(\ref{Eq:NonBD-uxi(tau)})--(\ref{Eq:NonBD-uyz(tau)}) are the main results of this appendix. They provide the leading order correction, i.e. up to the 1st order derivative of the pulse, to the non-BD subspace of the time evolution operator. Importantly, on top of the effective frequencies that appear in the BD subspace, characterized by $\omega_{\pm}$ and $\omega_{zi}$, there exists higher-frequency oscillation in the non-BD subspace that is set by qubit-qubit detuning $\Delta_{ct}$.      


\section{Off-resonantly driven transmon}
\label{App:OffResQu}

In this appendix, we consider an off-resonantly driven transmon qubit as a simpler model that still captures the main physics of off-resonant error on the control qubit. In comparison to the energy diagram in Fig.~\ref{fig:CRSchematicPlusEnergyDiagram}, this corresponds to one of the vertical ladders consisting of the control qubit states. First, in Sec.~\ref{SubApp:OffResQuOvInt}, we derive the leading order overlap integrals presented in Table~\ref{tab:OffResErrSummary}. Second, in Sec.~\ref{SubApp:OffResQuDRAG}, we derive leading order analytical DRAG solutions that minimize each specific error type. 

\subsection{Derivation of overlap integrals}
\label{SubApp:OffResQuOvInt}

In the rotating frame of the drive, the system and drive Hamiltonian for an off-resonantly driven transmon can be approximated using a Kerr model as
\begin{align}
&\HO_q \equiv \Delta_{cd} \hat{b}_c^{\dag}\hat{b}_c+\frac{\alpha_c}{2}\hat{b}_c^{\dag}\hat{b}_c^{\dag}\hat{b}_c\hat{b}_c  \;,
\label{Eq:OffResQu-Def of H0}\\
&\HO_{d} (t)  \equiv \frac{1}{2} \left[\Omega_c^*(t)\hat{b}_c + \Omega_c(t) \hat{b}_c^{\dag} \right]\;,
\label{Eq:OffResQu-Def of Hint(t)}
\end{align}
where $\Delta_{cd}\equiv \omega_c -\omega_d$ is the qubit-drive detuning. The interaction-frame Hamiltonian is then found as 
\begin{align}
\begin{split}
\HO_{I}(t) &\equiv e^{i \HO_q t}\HO_d(t)e^{-i \HO_q t}\\
& = \sum\limits_{n_c=1}^{d} \left[\frac{\sqrt{n_c}}{2}\Omega_c(t)e^{i \Delta_{nd} t}\ket{n_c}\bra{n_c-1}+\text{H.c.}\right] \;,
\end{split}	
\label{Eq:OffResQu-Def of HI(t)}
\end{align}
with level-dependent detuning $\Delta_{nd}\equiv \Delta_{cd}+(n_c-1)\alpha_c$ and level cut-off $d$. Assuming the detuning lies in the straddling regime, and in the drive range relevant to CR, the leakage to levels beyond the second excited state is typically negligible. Hence, a three-level model is sufficient for quantifying leading order transition probabilities $\ket{0_c}\rightarrow \ket{1_c}$, $\ket{0_c}\rightarrow \ket{2_c}$ and $\ket{1_c}\rightarrow \ket{2_c}$.

To solve for the time evolution operator, and the resulting overlap integrals, we follow the Magnus method up to the 2nd order, while the same result can also be obtained via time-dependent SWPT. A perturbative expansion of the time evolution operator $\hat{U}_I(\tau_p,0)\equiv \exp[-i\hat{K}(\tau_p,0)]$ in terms of the generator $\hat{K}(\tau_p,0)$ yields
\begin{subequations}
\begin{align}
\begin{split}
&\hat{U}_{I} (\tau_p,0) = \hat{I} - i\hat{K}_1(\tau_p,0)\\
&- i\hat{K}_2(\tau_p,0) -\frac{1}{2} \hat{K}_1^2(\tau_p,0)+O(\HO_I^3)\;,
\label{Eq:OffResQu-MagExp of Ui(t,t0)}
\end{split}
\end{align}
where $\hat{K}_1(\tau_p,0)$ and $\hat{K}_2(\tau_p,0)$ read \cite{Blanes_Pedagogical_2010}
\begin{align}
& \hat{K}_1(\tau_p,0)=\int_{0}^{\tau_p} dt' \HO_{I}(t')\;,
\label{Eq:OffResQu-Def of K1(tau,0)}\\
& \hat{K}_2(\tau_p,0)= -\frac{i}{2} \int_{0}^{\tau_p} dt' \int_{0}^{t'} dt'' [\HO_{I}(t'),\HO_{I}(t'')]\;.
\label{Eq:OffResQu-Def of K2(tau,0)}
\end{align}
\end{subequations}
In Eq.~(\ref{Eq:OffResQu-MagExp of Ui(t,t0)}), 1st order (single-photon) transitions occur via $-i\hat{K}_1(\tau_p,0)$, while 2nd order (two-photon) transitions occur via both $-i\hat{K}_2(\tau_p,0)$ and $-(1/2)\hat{K}_1^2(\tau_p,0)$. The two contributions up to the 2nd order, however, describe different physical processes: $-(1/2)\hat{K}_1^2(\tau_p,0)$ quantifies a two-photon spectral overlap as a product of individual single-photon overlaps, while $-i\hat{K}_2(\tau_p,0)$ quantifies a two-time (two-frequency) spectral overlap as shown in the following.

We begin by single-photon transitions. The transition probability of $\ket{0_c	}\rightarrow \ket{1_c}$ is found as
\begin{align}
\begin{split}
P_{\ket{0_c}\rightarrow\ket{1_c}} & = \left|\bra{1_c}\hat{U}_I(\tau_p,0)\ket{0_c}\right|^2\\
& = \left|-i \bra{1_c}\hat{K}_1(\tau_p,0)\ket{0_c}+O(\HO_I^3)\right|^2\\
&= \left|-i \int_{0}^{\tau_p} dt' \bra{1_c}\HO_I(t')\ket{0_c}+O(\HO_I^3)\right|^2
\end{split} 
\label{Eq:OffResQu-P0to1 generic}
\end{align}
Replacing $\bra{1_c}\HO_I(t')\ket{0_c}=(1/2)\Omega_c(t')e^{i\Delta_{cd} t'}$ from Eq.~(\ref{Eq:OffResQu-Def of HI(t)}) gives
\begin{align}
\begin{split}
P_{\ket{0_c}\rightarrow\ket{1_c}} &\approx \frac{1}{4}\left|\int_{0}^{\tau_p} dt' \Omega_c(t')e^{i\Delta_{cd} t'}\right|^2 \\
& = \frac{1}{4}\left|\int_{-\infty}^{+\infty} \frac{d\omega'}{2\pi} \frac{\tilde{\Omega}_c(\omega')}{\omega'+\Delta_{cd}}\left[e^{i(\omega'+\Delta_{cd})\tau_p}-1\right]\right|^2\;.
\end{split}
\label{Eq:OffResQu-P0to1 final} 
\end{align}
Equation~(\ref{Eq:OffResQu-P0to1 final}) is the leading order measure for non-BD error in the CR gate. It shows that the transition is run by a sideband photon as a result of spectral overlap between the pulse and qubit-drive detuning. Similarly, the transition probability of $\ket{1_c}\rightarrow \ket{2_c}$ reads
\begin{align}
\begin{split}
&P_{\ket{1_c}\rightarrow\ket{2_c}} \approx \frac{1}{2}\left|\int_{0}^{\tau_p} dt' \Omega_c(t')e^{i(\Delta_{cd}+\alpha_c) t'}\right|^2 \\
& = \frac{1}{2}\left|\int_{-\infty}^{+\infty} \frac{d\omega'}{2\pi} \frac{\tilde{\Omega}_c(\omega')}{\omega'+\Delta_{cd}+\alpha_c}\left[e^{i(\omega'+\Delta_{cd}+\alpha_c)\tau_p}-1\right]\right|^2 \;,
\end{split} 
\label{Eq:OffResQu-P1to2 final} 
\end{align}
where the distinct prefactor and transition frequency come from $\bra{2_c}\HO_I(t')\ket{1_c}=(\sqrt{2}/2)\Omega_c(t')e^{i(\Delta_{cd}+\alpha_c)t'}$.

Two-photon transition probability $\ket{0_c}\rightarrow \ket{2_c}$ is obtained as
\begin{align}
\begin{split}
P_{\ket{0_c}\rightarrow\ket{2_c}} &= \left|\bra{2_c}\hat{U}_I(\tau_p,0)\ket{0_c}\right|^2\\
& = \Big|-\frac{1}{2} \bra{2_c}\hat{K}_1^2(\tau_p,0)\ket{0_c}\\
&-i \bra{2_c}\hat{K}_2(\tau_p,0)\ket{0_c}+O(\HO_I^3)\Big|^2 \;,
\end{split}
\label{Eq:OffResQu-P0to2 generic}  
\end{align}
where contributions from $-\frac{1}{2}\hat{K}_1^2(\tau_p,0)$ and $-i\hat{K}_2(\tau_p,0)$ read
\begin{align}
\begin{split}
-\frac{1}{2} \bra{2_c}\hat{K}_1^2(\tau_p,0)\ket{0_c}=-\frac{\sqrt{2}}{8}\left[\int_0^{\tau_p} dt' \Omega_c(t')e^{i\Delta_{cd}t'}\right]\\
\times \left[\int_0^{\tau_p} dt' \Omega_c(t')e^{i(\Delta_{cd}+\alpha_c)t'}\right]\;,
\end{split}
\label{Eq:OffResQu-P0to2 K1 final}\\ 
\begin{split}
-i \bra{2_c}\hat{K}_2(\tau_p,0)\ket{0_c} = -\frac{\sqrt{2}}{8} \int_{0}^{\tau_p}dt'\int_{0}^{t'}dt'' \Omega_c(t')\Omega_c(t'') \\ 
\left[e^{i(\Delta_{ct}+\alpha_c)t'}e^{i\Delta_{ct}t''}-e^{i\Delta_{ct}t'}e^{i(\Delta_{cd}+\alpha_c)t''}\right]\;.
\end{split}
\label{Eq:OffResQu-P0to2 K2 final}
\end{align}
Equation~(\ref{Eq:OffResQu-P0to2 K1 final}) is a product of simultaneous overlaps with $\Delta_{cd}$ and $\Delta_{cd}+\alpha_c$. Depending on frequency allocation, either one or both overlaps are small and hence their product cannot grow too large. However, Eq.~(\ref{Eq:OffResQu-P0to2 K2 final}) contain higher order correlations leading to spectral overlap with $\ket{0_c}\rightarrow\ket{2_c}$ transition frequency $2\Delta_{cd}+\alpha_c$ as  
\begin{align}
\begin{split}
&P_{\ket{0_c}\rightarrow\ket{2_c}} \approx\frac{1}{32}\left|\int\limits_{-\infty}^{+\infty}\int\limits_{-\infty}^{+\infty} \frac{d\omega'}{2\pi} \frac{d\omega''}{2\pi} \alpha_c \tilde{\Omega}_c(\omega') \tilde{\Omega}_c(\omega'')\right.\\
&\left.\frac{\Big[e^{i(\omega'+\omega''+2\Delta_{ct}+\alpha_c)\tau_p}-1\Big]}{(\omega''+\Delta_{ct}+\alpha_c)(\omega''+\Delta_{ct})(\omega'+\omega''+2\Delta_{ct}+\alpha_c)}\right|^2 \;.
\end{split}
\label{Eq:OffResQu-P0to2 FourierRep}
\end{align}
Equation~(\ref{Eq:OffResQu-P0to2 FourierRep}) is an approximate form for the frequency-domain representation of Eq.~(\ref{Eq:OffResQu-P0to2 K2 final}). Here, where we kept only the dominant two-photon contribution in which the transition is excited by \textit{two} sideband photons with total energy $\omega'+\omega''=-(2\Delta_{cd}+\alpha_c)$. The overlap grows when the two-photon gap is small, i.e. close to the collision at $\Delta_{cd}=-\alpha_c/2$.  

\subsection{Derivation of DRAG conditions} 
\label{SubApp:OffResQuDRAG}

We next derive leading order DRAG solutions to suppress single-photon transitions discussed in Eqs.~(\ref{Eq:OffResQu-P0to1 final}) and~(\ref{Eq:OffResQu-P1to2 final}). The derivation here is slightly distinct compared to Refs.~\cite{Motzoi_Simple_2009, Gambetta_Analytic_2011} and done in two consecutive steps: we first perform perturbation in drive amplitude $\Omega_c(t)$ and then an adiabatic expansion in $\frac{1}{\Delta} \frac{d}{dt}$ allowing for more flexibility. Nevertheless, we recover the solution in Ref.~\cite{Motzoi_Simple_2009} as a special case. 

To this aim, we consider the following leading order $Y$-DRAG Ansatz 
\begin{align}
\Omega_c(t)=\Omega_{\text{SG}}(t)+\frac{i}{\Delta_D}\dot{\Omega}_{\text{SG}}(t)\;,
\label{Eq:OffResQu-DRAG Ansatz}
\end{align}
where the DRAG coefficient $\Delta_D$ needs to be determined such that a specific off-resonant overlap is suppressed. Here, the main pulse is taken as square Gaussian, but the DRAG condition is in principle independent of this choice. Performing an adiabtic expansion on $\ket{0_c}\rightarrow \ket{1_c}$ probability in Eq.~(\ref{Eq:OffResQu-P0to1 final}) results in
\begin{align}
\begin{split}
&P_{\ket{0_c}\rightarrow\ket{1_c}} \approx \frac{1}{4} \left|\int_{0}^{\tau_p} dt'\Omega_c(t') e^{i \Delta_{cd} t'}\right|^2\\
&=\frac{1}{4} \left|\Big\{\sum\limits_{n=0}^{\infty}\Big[\frac{1}{\Delta_{cd}}\Big(\frac{i}{\Delta_{cd}}\frac{d}{dt'}\Big)^n\Omega_c(t')\Big]e^{i\Delta_{cd} t'}\Big\} \Big|_{0}^{\tau_p}\right|^2\;.
\end{split}
\label{Eq:OffResQu-P0To1 AdiabExp 1}
\end{align}
We then substitute the DRAG Ansatz~(\ref{Eq:OffResQu-DRAG Ansatz}) into Eq.~(\ref{Eq:OffResQu-P0To1 AdiabExp 1}) and read off contributions at $t=0,\tau_p$ as a measure for off-resonant error. In terms of normalized DRAG parameter $\lambda_{01}\equiv \Delta_{cd}/\Delta_D$, and given that $\Omega_{\text{SG}}(0)=\Omega_{\text{SG}}(\tau_p)=0$, we find
\begin{widetext}
\begin{align}
\begin{split}
P_{\ket{0_c}\rightarrow\ket{1_c}}=\frac{(1+\lambda_{01})^2}{4\Delta_{cd}^4}\left[\dot{\Omega}_{\text{SG}}^2(\tau_p)+\dot{\Omega}_{\text{SG}}^2(0)-2\cos(\Delta
_{cd}\tau_p)\dot{\Omega}_{\text{SG}}(\tau_p)\dot{\Omega}_{\text{SG}}(0)\right]\\
+\frac{(1+\lambda_{01})^2}{4\Delta_{cd}^6}\left[\ddot{\Omega}_{\text{SG}}^2(\tau_p)+\ddot{\Omega}_{\text{SG}}^2(0)-2\cos(\Delta
_{cd}\tau_p)\ddot{\Omega}_{\text{SG}}(\tau_p)\ddot{\Omega}_{\text{SG}}(0)\right]\\
+\frac{\lambda_{01}^2}{4\Delta_{cd}^8}\left[\dddot{\Omega}_{\text{SG}}^2(\tau_p)+\dddot{\Omega}_{\text{SG}}^2(0)-2\cos(\Delta
_{cd}\tau_p)\dddot{\Omega}_{\text{SG}}(\tau_p)\dddot{\Omega}_{\text{SG}}(0)\right]\\
-\frac{\lambda_{01}(1+\lambda_{01})}{2\Delta_{cd}^6}\Big[\dddot{\Omega}_{\text{SG}}(\tau_p)\dot{\Omega}_{\text{SG}}(\tau_p)+\dddot{\Omega}_{\text{SG}}(0)\dot{\Omega}_{\text{SG}}(0)\\
-\cos(\Delta_{cd}\tau_p)\dddot{\Omega}_{\text{SG}}(\tau_p)\dot{\Omega}_{\text{SG}}(0)-\cos(\Delta_{cd}\tau_p)\dddot{\Omega}_{\text{SG}}(0)\dot{\Omega}_{\text{SG}}(\tau_p)\Big] 	
+O\left(\frac{d^4}{dt^4}\Omega_{\text{SG}}(t)\Big|_{t=0,\tau_p}\right)\;.	
\end{split}
\label{Eq:OffResQu-P0To1 AdiabExp 2}
\end{align}
\end{widetext}
Based on Eq.~(\ref{Eq:OffResQu-P0To1 AdiabExp 2}), up to the leading order, the optimal DRAG parameter $\lambda_{01}$ is determined as the roots of a 2nd order polynomal, whose coefficients are generally determined by the pulse spectrum (derivatives), gate time $\tau_p$ and the transition frequency $\Delta_{cd}$. A \textit{special} solution, however, is found as $\lambda_{01}=-1$, i.e. $\Delta_D=-\Delta_{ct}$, which sets the 1st, 2nd and the 4th term in Eq.~(\ref{Eq:OffResQu-P0To1 AdiabExp 2}) to zero resulting in residual error in terms of $\dddot{\Omega}_{\text{SG}}(t)$ as
\begin{align}	
\frac{\left[\dddot{\Omega}_{\text{SG}}^2(\tau_p)+\dddot{\Omega}_{\text{SG}}^2(0)-2 \cos(\Delta_{cd}\tau_p)\dddot{\Omega}_{\text{SG}}(\tau_p)\dddot{\Omega}_{\text{SG}}(0)\right]}{4\Delta_{cd}^8} \;.
\label{Eq:OffResQu-P0To1 AdiabExp 3}
\end{align} 
This DRAG choice corresponds to $X$ and $Y$ control pulses $\Omega_{cx}(t)=\Omega_{\text{SG}}(t)$ and $\Omega_{cy}(t)=-(1/\Delta_{cd})\dot{\Omega}_{\text{SG}}(t)$.

Adiabatic expansion of the $\ket{1_c}\rightarrow\ket{2_c}$ transition probability has a similar form as
\begin{align}
\begin{split}
P_{\ket{1_c}\rightarrow\ket{2_c}} \approx \frac{1}{2} \left|\int_{0}^{\tau_p} dt'\Omega_c(t') e^{i (\Delta_{cd}+\alpha_c) t'}\right|^2\\
=\frac{1}{2} \left|\Big\{\sum\limits_{n=0}^{\infty}\Big[\frac{1}{\Delta_{cd}+\alpha_c}\Big(\frac{i}{\Delta_{cd}+\alpha_c}\frac{d}{dt'}\Big)^n\Omega_c(t')\Big]\right.\\
\times e^{i (\Delta_{cd}+\alpha_c) t'}\Big\} \Big|_{0}^{\tau_p}\Bigg|^2\;,
\end{split}
\label{Eq:OffResQu-P1To2 AdiabExp 1}
\end{align}
where the transition frequency is replaced by $\Delta_{cd}+\alpha_c$. In terms of $\lambda_{12}\equiv (\Delta_{cd}+\alpha_c)/\Delta_D$, the derivative expansion takes the same form as in Eq.~(\ref{Eq:OffResQu-P0To1 AdiabExp 2}) and a possible DRAG solution is $\Omega_{cx}(t)=\Omega_{\text{SG}}(t)$ and $\Omega_{cy}(t)=-1/(\Delta_{cd}+\alpha_c)\dot{\Omega}_{\text{SG}}(t)$. In the resonant scenario where $\Delta_{cd}=0$, this solution recovers that of Ref.~\cite{Motzoi_Simple_2009} for single qubit gates.

We note that the same method of adiabatic expansion can also be applied to the $\ket{0_c}\rightarrow \ket{2_c}$ probability in Eqs.~(\ref{Eq:OffResQu-P0to2 generic})--(\ref{Eq:OffResQu-P0to2 K2 final}). However, the computation is much more involved, and the leading order DRAG condition is determined as the roots of a 4th order polynomial in $1/\Delta_D$. Therefore, we resort to numerical optimization as discussed in Sec.~\ref{Sec:DRAG} and Fig.~\ref{fig:DRAGSweep}. 


\bibliographystyle{unsrt}
\bibliography{QuantumBibliography}
\end{document}